\documentclass[twocolumn,showpacs,longbibliography,nofootinbib,superscriptaddress,prb]{revtex4-2}
\usepackage{graphicx}
\usepackage{textcomp}
\usepackage{amsmath,amssymb}
\usepackage{color}
\usepackage{soul}
\usepackage{mathtools}
\usepackage{bbold}
\usepackage{multirow}
\newcommand{\newc}{\newcommand}
\newc{\beq}{\begin{equation}}
\newc{\eeq}{\end{equation}}
\newc{\kt}{\rangle}
\newc{\br}{\langle}
\newc{\beqa}{\begin{eqnarray}}
\newc{\eeqa}{\end{eqnarray}}

\newc{\Tr}{\mbox{Tr}}
\newc{\ovl}{\overline}
\usepackage[colorlinks=true, urlcolor=blue, citecolor=blue, linkcolor=blue]{hyperref}
\begin{document}
\title{Out-of-time-order correlators of nonlocal block-spin and random observables in integrable and nonintegrable spin chains}
\author{Rohit Kumar Shukla}
\email[]{rohitkrshukla.rs.phy17@itbhu.ac.in}
\affiliation{Department of Physics, Indian Institute of Technology (Banaras Hindu University), Varanasi - 221005, India}
\author{Arul Lakshminarayan}
\email[]{arul@physics.iitm.ac.in}
\affiliation{Department of Physics, Indian Institute of Technology Madras, Chennai - 600036, India}
\author{Sunil Kumar Mishra}
\email[]{sunilkm.app@iitbhu.ac.in}
\affiliation{Department of Physics, Indian Institute of Technology (Banaras Hindu University), Varanasi - 221005, India}
\date{\today}

\begin{abstract}
Out-of-time-order correlators (OTOC) in the Ising Floquet system, that can be both integrable and nonintegrable is studied. Instead of localized spin observables, we study contiguous symmetric blocks of spins or random operators localized on these blocks as observables. We find only power-law growth of OTOC in both integrable and nonintegrable regimes. In the non-integrable regime, beyond the scrambling time, there is an exponential saturation of the OTOC to values consistent with random matrix theory. This motivates the use of ``pre-scrambled" random block operators as observables. A pure exponential saturation of OTOC in both integrable and nonintegrable system is observed, without a scrambling phase. Averaging over random observables from the Gaussian unitary ensemble, the OTOC is found to be exactly  same as the operator entanglement entropy, whose exponential saturation has been observed in previous studies of such spin-chains.
\end{abstract}

\maketitle
\section{Introduction}
\label{Introduction}
Periodically driven Floquet systems have been extensively studied
in the recent past in both classical and quantum system. A popular set of models are driven by fields applied in the form of kicks \cite{d2014long, naik2019controlled, shukla2021,Mishra2015}, as analytical forms of the time evolution operator are easy to find.
One textbook example is the kicked-rotor model of a particle moving on a ring \cite{reichl2004transition}. These models show interesting behavior displaying transition from integrability to chaos, dynamical Anderson localization \cite{chirikov1981dynamical,fishman1982chaos,reichl2004transition}, and dynamical stabilization \cite{kapitza1965dynamical,broer2004parametrically}. These systems are of interest in  both classical
as well as quantum systems. Such periodic forcing has been realised in experiments to study various phenomena \cite{wintersperger2020realization,franca2021simulating,zhang2017observation,choi2017observation,Santhanam2022}.
\par
In contrast to the kicked rotor,  the Ising model with time-periodic transverse and longitudinal magnetic fields is an example of a many-body Floquet system of current interest \cite{gritsev2017,lakshminarayan2005,naik2019controlled,shukla2021}.
 Absence of a transverse component renders the system trivially integrable. Presence of both a longitudinal and transverse magnetic component makes this system nonintegrable. However, in the absence of longitudinal field, the system is rendered integrable as a system of noninteracting fermions.   
These systems have been studied using sudden quenches \cite{Polkovnikov2011} and slow annealing \cite{Santoro2002}. In the quenched case, the system is out of equilibrium and leads to interesting dynamics of the observables,  and has drawn considerable attention in the last decade with significant theoretical and experimental observations \cite{Russomanno2012,Russomanno2013,mishra2014resonance}. 

\par
A typical way to distinguish between integrable, non-integrable and near-integrable regimes has been to use spectral properties and random matrix theory. This mostly leaves aside the question of dynamics. However, a quantity that has been extensively used recently to distinguish the chaotic and integrable dynamics, is the out-of-time-order correlator (OTOC) \cite{ garcia2018,rozenbaum2020,yan2019,rozenbaum2017,lee2019,rozenbaum2019}. In classical physics, one hallmark of chaos is that a small difference in the initial condition results in the exponential deviation of the trajectory, which is responsible for the so-called ``butterfly effect" \cite{gu2016,bilitewski2018temperature,das2018light}. Classical Hamiltonian systems can have such pure deterministic chaos which is used in the quantum domain for the study of quantum chaos \cite{gutzwiller2013,haake1991}. It has been proposed that quantum chaos be characterized by the growth rate of OTOC \cite{Maldacena2016}, an  exponential growth defining a quantum Lyapunov exponent. 
\par
Spin systems have been a playground for understanding  many-body physics in general and growth of OTOCs in particular  \cite{lin2018out,xu2020accessing,xu2019locality,kukuljan2017weak,Fortes2019,Craps2020,roy2021entanglement,yan2019similar,bao2020out,dora2017out,Riddell2019,lee2019typical}.  Growth of OTOC is discussed in systems such as Luttinger liquids \cite{dora2017out}, XY model \cite{bao2020out}, Sachdev-Ye-Kitaev (SYK) model \cite{Fu2016} , Heisenberg XXZ model and  Aubry–Andr\'e–Harper model \cite{Riddell2019,lee2019typical}.
Lin and Motrunich \cite{lin2018out} calculated OTOC for single spin observables in the integrable transverse field Ising model, and observed power-law growth, with the power varying with the separation between the localized spins.

Fortes {\it et. al} \cite{Fortes2019} studied OTOCs in the time independent Ising model with tilted magnetic fields, perturbed XXZ model, and Heisenberg spin model with random magnetic fields. In all these models with single-spin  observables, only power-law growth has been reported despite the presence of quantum chaos. 
OTOCs in integrable and nonintegrable Floquet Ising  models were studied by Kukuljan {\it et. al.} \cite{kukuljan2017weak} using  extensive observables. In one dimension case, the growth of OTOC density was still found to be linear in time.

The cases where exponential growth
has been definitely reported involve semiclassical models such as the quantum kicked rotor \cite{rozenbaum2017lyapunov}, coupled kicked rotors \cite{prakash2020scrambling,Santhanam2022}, the kicked top which may be considered to be a transverse field kicked Ising model but with the interactions being all-to-all \cite{yin2021quantum,sreeram2021out}, the bakers map \cite{lakshminarayan2019out}, and so on. Our motivation herein is to allow for a large Hilbert space for the observables, which are restricted to blocks of spins. We may consider the spin chain as a bipartite chaotic system each consisting of $N/2$ spins, to explore the possibility of exponential growth. We will see that such spin-1/2 nonintegrable models, even for block operators have only power-law OTOC growth, implying that their quantum Lyapunov exponents are $0$. 

\par
In nonintegrable systems including spin chains such as studied here the long time  saturation value of the OTOC is consistent with an estimate from random matrix theory. The approach of the OTOC to the saturation value was found to be at an exponential rate in weakly interacting bipartite chaotic system \cite{prakash2020scrambling}. Exponential approach to saturation was also found in a semiclassical theory of OTOC \cite{rammensee2018many}.
We find such an exponential approach to the random matrix value in spin chains with block observables for the nonintegrable cases. 

To understand the exponential approach, we consider the case when the block operators are random. Averaging over random unitary operators in bipartite system, the OTOC has been shown to be exactly the operator entanglement of the propagator \cite{anand2021brotocs}. We show this is also the case with random Hermitian observables, drawn from the Gaussian Unitary Ensemble (GUE). 

Thus the exponential saturation of the OTOC is qualitatively consistent with the behavior previously observed for the operator entanglement growth of the propagator \cite{Pal2018}.  

According to the BGS conjecture \cite{Bohigas1984}, the spectral properties of the quantum analogue of a chaotic classical system will follow Wigner-Dyson statistics  unlike the quantum analogue of a integrable classical system following Poisson distribution.  Thus, the spectral statistics of spacing between the consecutive energy levels of a quantum system works as a tool to differentiate a chaotic system from an integrable one    \cite{craps2020lyapunov,Pal2018,karthik2007entanglement,Ray2018,chen2018measuring,ray2018signature,mehta1991theory,averbukh2001angular}.
\par
This manuscript is organised as follows. In \ref{model}, we will discuss the Floquet map with and without longitudinal fields. In \ref{OTOC}, we will define the OTOC for the block spin operators. In \ref{avg}, we will discuss the relation of OTOC with operator entanglement entropy (OPEE).
In the \ref{LSD}, we will elaborate the nearest-neighbour spacing distribution (NNSD) and its behavior in the integrable and nonintegrable cases. we will elaborate the behavior of OTOC and NNSD in \ref{CF}, for the constant-field Flqouet system and in the  \ref{special_CF}, a special case of constant field Flquet system. Finally in \ref{conclusion}, we will conclude the results of the manuscript.

\maketitle
\section{The spin model and background}
\subsection{The spin model}
\label{model}

Consider a periodically driven Ising spin system with the Hamiltonian
\begin{equation}
\label{Hxz}
\hat H(t)=J_{x} \hat H_{xx}+h_{x} \hat H_x+h_{z}\sum_{n=-\infty}^{\infty}\delta\Big(n-\frac{t}{\tau}\Big) \hat H_z.
\end{equation}
Here $\hat H_{xx}=\sum_{l=1}^{N-1}\hat \sigma_l^x\hat \sigma_{i+1}^x $ is the nearest-neighbor Ising interaction term,  $ \hat H_x=\sum_{l=1}^{N}\hat \sigma_l^x$ and $\hat H_z=\sum_{l=1}^{N}\hat \sigma_l^z$. The interaction strength is $J_x$, the continuous and constant longitudinal magnetic field in $x$-direction is given by $h_{x}$ and the transverse magnetic field in the $z$-direction, which is applied in the form of delta pulses at regular interval $\tau$ is $h_{z}$.
 
The Floquet operator is the propagator connecting states across one time period $\tau$. Denoting this as $\mathcal{\hat U}_x$, we have (with $\hbar=1$)
\begin{equation}
\label{Ux}
 \mathcal{\hat U}_x =\exp\left[-i\tau(J_x \hat H_{xx}+h_{x}\hat H_{x})\right] \exp\left(-i \tau h_{z} \hat H_{z}\right), 
\end{equation}
and will be referred to as $``\mathcal{\hat U}_x$ systems" below,  
When the longitudinal field is absent the model is solvable by 
the Jordan–Wigner transformation and renders the system as one of noninteracting fermions. In the presence of the longitudinal field these fermions are interacting and there is evidence that there is a transition to quantum chaos
 \cite{prosen2004ruelle,prosen2000exact,prosen2002general,lakshminarayan2005multipartite,else2016floquet}. The Floquet map of integrable model is a special case of Eq. (\ref{Ux}) with $h_{x}=0$ will be referred to as the $\mathcal{\hat U}_0$ system below.

\subsection{Out-of-time-order correlation and block operators}
\label{OTOC}

Dynamics of quantum systems lead to the spreading of initially localized operators under the unitary time evolution.  Let the discrete time evolution of operator $\hat W \equiv \hat W(0)$ be $\hat W(n)=\hat U(n)^{\dagger}\hat W(0)\hat U(n)$, where $\hat U(n)$ is time$-n$ propagator. For example if the time evolution is governed by Eq. (\ref{Ux}), $\hat U(n)= \mathcal{\hat U}_x^n$.
If $\hat V$ and $\hat W$ are two Hermitian operators that are localized on different sets of spins (say $A$ and $B$), we consider as the out-of-time-order correlation (OTOC)  \cite{larkin1968zh,shenker2014black,shenker2014multiple,almheiri2013apologia,shenker2015stringy,roberts2015diagnosing,maldacena2016bound,stanford2016many}:
\begin{eqnarray}
C(n)=-\frac{1}{2 \, d_A d_B}\mbox{Tr} \left([\hat W(n), \hat V]^2\right),
\label{Cn1}
\end{eqnarray}
where $d_A$ and $d_B$ are dimensions of the subspaces, and  $d_A=d_B=2^{N/2}$ as we consider only the case of equal blocks. The OTOC $C(n)$ is clearly a measure of the noncommutativity of these two operators, via its norm.

This separates as $C(n)$ =$C_2(n)-C_4(n)$, where $C_2(n)$ and $C_4(n)$ are two-point and four-point correlations respectively:
\begin{eqnarray}
\label{cn2}
C_2(n)=\frac{1}{d_A d_B}\mbox{Tr}(\hat W^2(n) \hat{V}^2),\\ 
\label{cn4}
C_4(n)=\frac{1}{d_A d_B}\mbox{Tr}(\hat W(n)\hat V\hat W(n)\hat V).
\end{eqnarray}
These are infinite temperature quantities and involves the entire spectrum of $2^{N}$ states. We will use the trick of evaluating this by employing 
Haar random states of $2^N$ dimensions to evaluate expectation values, that is 
$\mbox{Tr}(\hat A)/2^N \approx \left \langle \Psi_R|\hat A|\Psi_R\right \rangle$
were $|\Psi_R\rangle$ is such a state. Averages over a few random states are used.

\begin{figure}[hbt!]
      \centering
       \includegraphics[width=\linewidth, height=.30\linewidth]{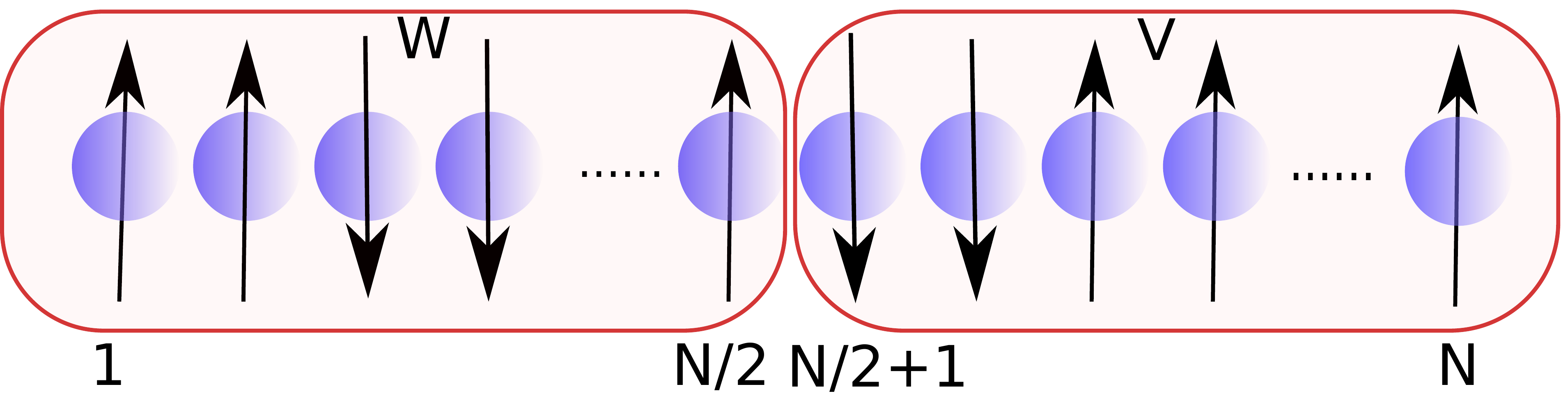}
       \caption{Schematics of SBOs defined in Eq. (\ref{Block}).  Even $N$ is considered and halved into subsystems $W$ and $V$.}
       \label{block_operator} 
\end{figure}

Almost all studies of OTOC in such spin models thus far concentrate on operators that are localized on single spins, in contrast we consider operators $\hat V$ and $\hat W$ initially isolated on the first and second block of spins, see Fig. \ref{block_operator}, referred to here as spin-block-operators (SBOs): 
  \begin{eqnarray}
  \label{Block}
\hat W=\frac{2}{N} \sum_{l=1}^{\frac{N}{2}}\hat \sigma_l^x \quad {\rm and} \quad \hat V=\frac{2}{N} \sum_{l=\frac{N}{2}+1}^{N}\hat \sigma_l^x.
\end{eqnarray}
Note that the behaviour of these OTOC are genuinely different and do not follow from a knowledge of the single site OTOCs involving correlations such as $\langle \hat{\sigma}^x_{l_1} \, \hat{\sigma}^x_{l_2}(n) \,\hat{\sigma}^x_{l_3}\, \hat{\sigma}^x_{l_4}(n) \rangle$ for general values of $l_i$.
For $n>0$, $\hat W(n)$ is no longer confined to the first $N/2$ spins, and the OTOC becomes nonzero.  

\subsection{Average and asymptotic OTOC values}
\label{avg}
 
As $\hat{V}$ and $\hat{W}$ are block restricted sums of spin operators, $\hat{V}+\hat{W}$ is the total spin in the ${x}$ direction and appears as a term in the Hamiltonian. Thus these are special operators with dynamical significance, as would be natural to assume. In contrast if they are random operators on the space of 
$N/2$ spins, the OTOC behaves quite differently till possibly the
scrambling time. Beyond the scrambling time, we may expect that the local operators have largely become random if there is nonintegrability and quantum chaos. Thus, it is of interest to compare the behaviour of random operator OTOC with non-random ones: to separate the effects of dynamics and scrambling. In a semiclassical model of weakly coupled chaotic systems, it was noted that the post-scrambling time OTOC of non-random operators did behave as that of ``pre-scrambled" random operators \cite{prakash2020scrambling}. We find some similartie in the case of spin chains, but also interesting differences. 

In the case of random operators for $\hat V$ and $\hat W$, ergodicity maybe expected and hence an average over them is done. It has been observed \cite{styliaris2021information} that if these operators are random unitaries chosen uniformly (Haar measure, circular unitary ensemble, CUE), the average OTOC is remarkably related to the operator entanglement. As we are using Hermitian operators, we average over random Hermitian ensembles for which we naturally choose the GUE, and the result is identical.

Let there be a bipartite space $\mathcal{\hat H}_{A}\otimes \mathcal{\hat H}_B$, such as the space of the first and second $N/2$ spins in the chain. The Schmidt decomposition of the unitary propagagtor on this bipartition is of the form 
\[ \hat U(n)=2^{N/2}\sum_{i=1}^{2^N}\sqrt{\lambda_i(n)}\,\hat A_i(n) \otimes \hat B_i(n).\]
Here $\hat A_i(n)$ and $\hat B_i(n)$ are orthonormal operators on individual spaces $\mathcal{\hat H}_{A,B}$,  satisfying, $\Tr (\hat A_i(n)^{\dagger} \hat A_j(n))= \Tr(\hat B_i(n)^{\dagger} \hat B_j(n))=\delta_{ij}$. The numbers $\lambda_i(n)>0$ and satisfy the condition $\sum_i\lambda_i(n)=1$ which is a consequence of the unitarity of $\hat U(n)$.
     
 Operator entanglement entropy (OPEE) is used for the measure of entanglement \cite{Pal2018,styliaris2021information,wang2002quantum,wang2004entanglement} and defined via the linear entropy as
 \begin{eqnarray}
 E_l[\hat U(n)]=1-\sum_{i=1}^{2^{N}}\lambda^2_i(n).
  \end{eqnarray}
This vanishes if and only if $\hat U(n)$ is of product form and is maximum when all $\lambda_i(n)=2^{-N}$ and the OPEE is equal to $1-2^{-N}$.

Let an element of the GUE be $\hat W=(\hat M+\hat M^{\dagger})/2$, where $\hat M$ is a $d$ dimensional matrix whose entries are such that its real and imaginary parts are zero centered, unit variance,  independent normal random numbers, the Ginibre ensemble. It is straightforward to see that $\ovl{\hat W^2}=d\, \hat I_d$, where $\hat I_d$ is the $d$ dimensional identity matrix, and the overline indicates the GUE average. The average of $C_2(n)$ is then 
\beq
\ovl{C_2(n)}^{\hat W,\hat V}=\frac{1}{d^2}\ovl{\Tr \left(\hat U(n)^{\dagger} \hat W^2\hat U(n)\hat \hat V^2\right) }^{\hat W,\hat V}=d^2,
\eeq
where $\hat V$ is also a GUE realization independent of $\hat W$. 

To evaluate the 4-point function $C_4(n)$, we need to use the standard ploy of doubling the space: $\Tr(\hat A^2)=\Tr((\hat A \otimes \hat A)\;\hat S)$ where $\hat S$ swaps the original and ancilla spaces. With $\hat A=\hat W(n)\hat V$ The only relevant average needed is 
\beq
\ovl{\hat W \otimes \hat W}^{\hat W}= \hat S,
\eeq 
and it follows using identities known for the operator entanglement \cite{anand2021brotocs,styliaris2021information} that $\ovl{C_4(n)}^{\hat W,\hat V}= d^2[1-E_l(\hat U(n))]$ and hence the OTOC averaged over the observables is 
\beq
\label{OPEE_OTOC}
\ovl{C(n)}^{\hat W,\hat V}=d^2 E_l[\hat U(n)].
\eeq
Thus the observable averaged OTOC is identical to the OPEE. Based on ergodicity, the case of a single random realization may then be expected to be represented by this average.

In the asymptotic limit of large times, if the dynamics is chaotic, we may expect that $\hat U(n)$ is a complex operator on the whole Hilbert space and treat it as being sampled according to the random CUE of size $2^{N}$, while keeping the $\hat W$ and $\hat V$ as fixed or non-random operators. The averaged quantities for traceless operators $\hat V$ and $\hat W$ are (see Appendix \ref{appendix1} for details) 
\begin{subequations}
\begin{align}
\overline{C_2(n)}^{U}&= \frac{1}{d^2} \Tr (\hat W^2) \Tr (\hat V^2)\\
\overline{C_4(n)}^{U}&=\frac{-1}{d^2(d^2-1)}\Tr (\hat W^2) \Tr (\hat V^2)\\
\overline{C(n)}^{U}&=\frac{1}{d^2-1} \Tr (\hat W^2) \Tr (\hat V^2).
\end{align}
\end{subequations}
For the $\hat W$ and $\hat V$ in Eq.~(\ref{Block}) the asymptotic value of the OTOC, ignoring the $C_4$ value, which is of lower order in the Hilbert space dimension, 
is this average and denoted below as
\beq
C(\infty)=4/N^2.
\eeq
For the GUE random $\hat V$ and $\hat W$ used above $\overline{\Tr \hat W^2}=d^2$ and hence in this case $C(\infty)=d^2=2^{2N}$ for large $d$. We will always study scaled OTOC, dividing by the relevant $C(\infty)$, thus for the random operator case, the averaged and scaled OTOC is exactly the OPEE $E_l[\hat U(n)]$.

\subsection{Nearest-neighbour spacing distribution}
 \label{LSD}
\par
 Spectral statistics of the spacing between consecutive energy levels is used to differentiate the chaotic and integrable regimes. In order to calculate the NNSD, first we need to identify the symmetries of the Hamiltonian. Next, the Hamiltonian is block diagonalized in the symmetry sectors. Our system with open boundary condition has a ``bit-reversal” symmetry at all the Floquet periods. This bit-reversal symmetry is due to the fact that the field and interaction do not distinguish the spins by interchanging the spins at the sites $i$ and $N-i+1$ for all $i=1,\cdots,N$. Let us consider $\hat B$ a bit-reversal operator given by
\begin{eqnarray}
\hat B|\mathtt{s}_1,\mathtt{s}_2,\cdots,\mathtt{s}_N\rangle=|\mathtt{s}_N,\cdots,\mathtt{s}_2,\mathtt{s}_1\rangle,\hspace{.5cm} [\hat U,\hat B]=0,\nonumber\\
\end{eqnarray}
where $|\mathtt{s}_i\rangle$ is any single-particle basis state in standard $(\mathtt{s}_z)$ basis. We divide whole basis sets into two groups of basis states: one with the palindrome in which there is no change in the state after applying the operator $\hat B$ {\it{i.e.}}, $\hat B|\mathtt{s}_1,\mathtt{s}_2,\cdots,\mathtt{s}_N\rangle=|\mathtt{s}_1,\mathtt{s}_2,\cdots,\mathtt{s}_N\rangle$. The other one with the non-palindrome in which states get reflected after applying the operator $\hat B$ {\it{i.e.}} $\hat B|\mathtt{s}_1,\mathtt{s}_2,\cdots,\mathtt{s}_N\rangle=|\mathtt{s}_N,\cdots,\mathtt{s}_2,\mathtt{s}_1\rangle$.  Since $\hat B^2=1$, the eigenvalues of $\hat B$ are $\pm1$. The eigenstates can be classified as odd or even state under bit-reversal. All the palindromes define even states, however all the non-palindromes correspond to one even and one odd state. Sum and difference of the non-palindrome and its reflection generate these even and odd states. 

We study the shape of distribution by using the NNSD which may be used as an indicator of quantum chaos and nontrivial integrable models. In NNSD, strongly chaotic points are those where the unfolded level-spacings are well described by the Wigner distribution \cite{Luca2016,mehta1991theory,averbukh2001angular} which is given as
\begin{eqnarray}
P_W(s)=\frac{\pi s}{2}e^{-\pi s^2/4},
\end{eqnarray}
 where, $s$ is drawn from the ensemble of consecutive energy level separation. On the other hand, nontrivial integrable models are those where the unfolded NNSD follows Poisson statistics,
\begin{eqnarray}
P_P(s)=e^{-s}.
\end{eqnarray}

\begin{figure*}[hbt!]
      \centering
       \includegraphics[width=.45\linewidth, height=.30\linewidth]{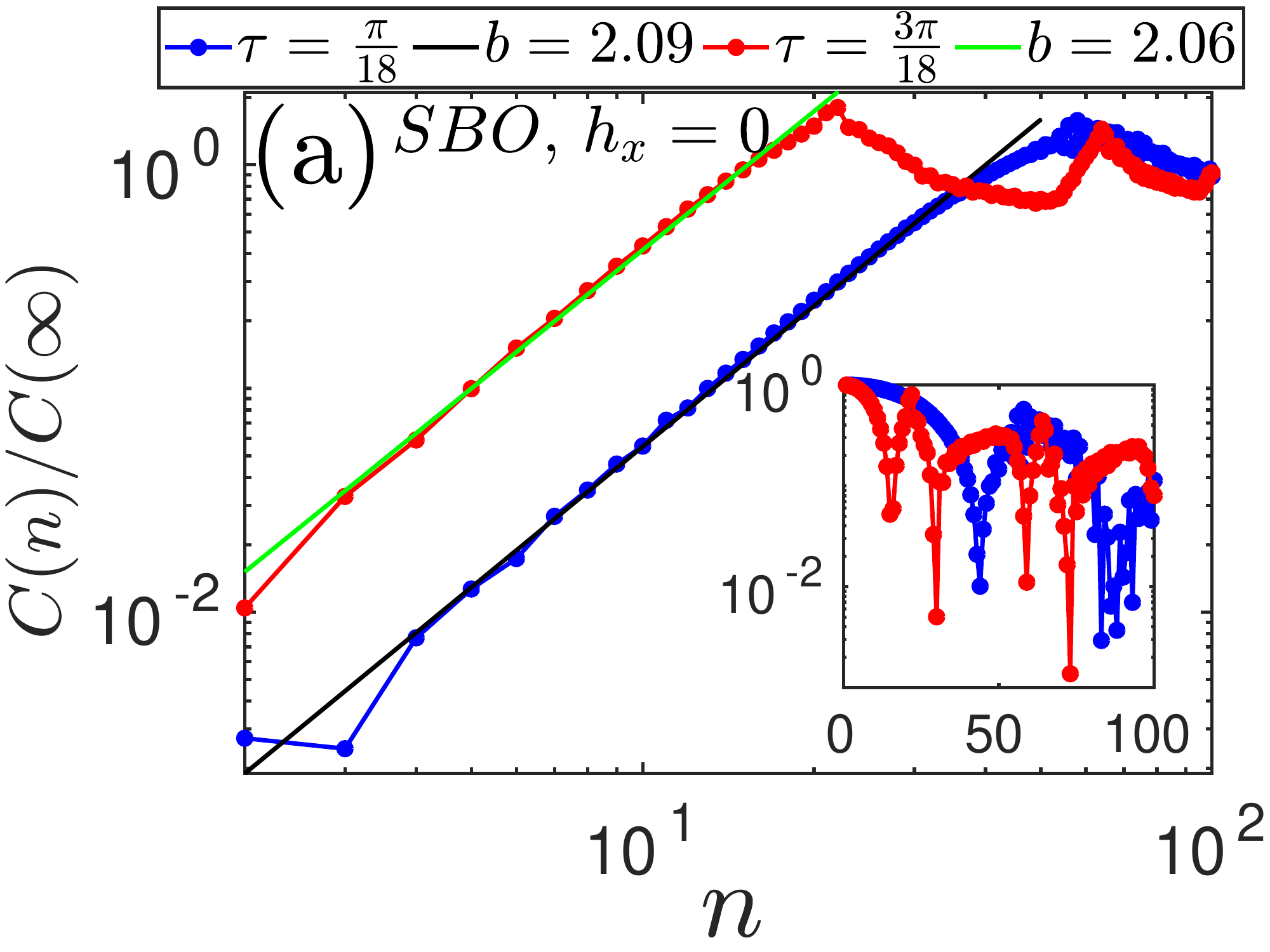}
       \includegraphics[width=.45\linewidth,height=.30\linewidth]{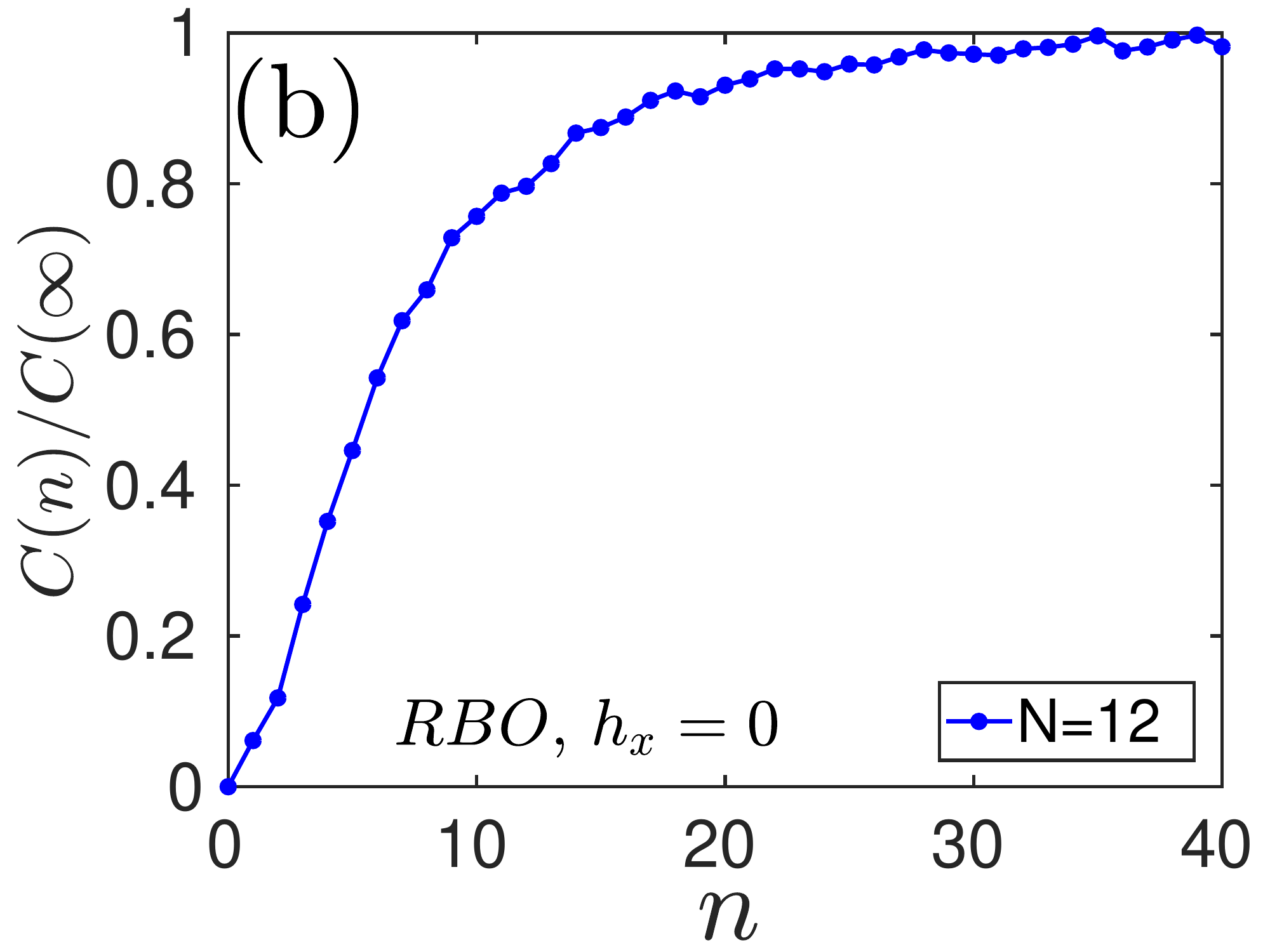}
       \includegraphics[width=.45\linewidth,height=.30\linewidth]{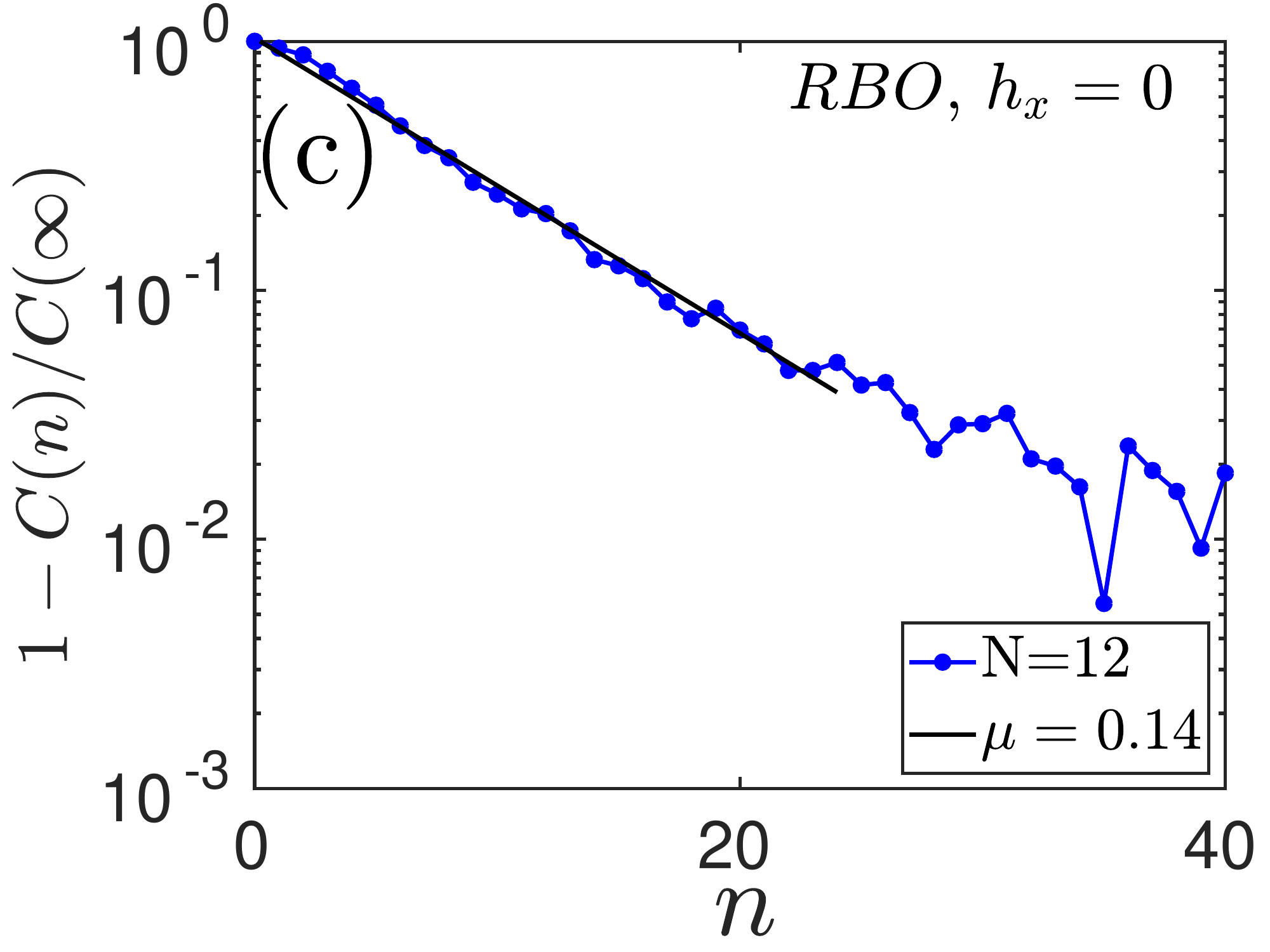}
       \includegraphics[width=.45\linewidth, height=.30\linewidth]{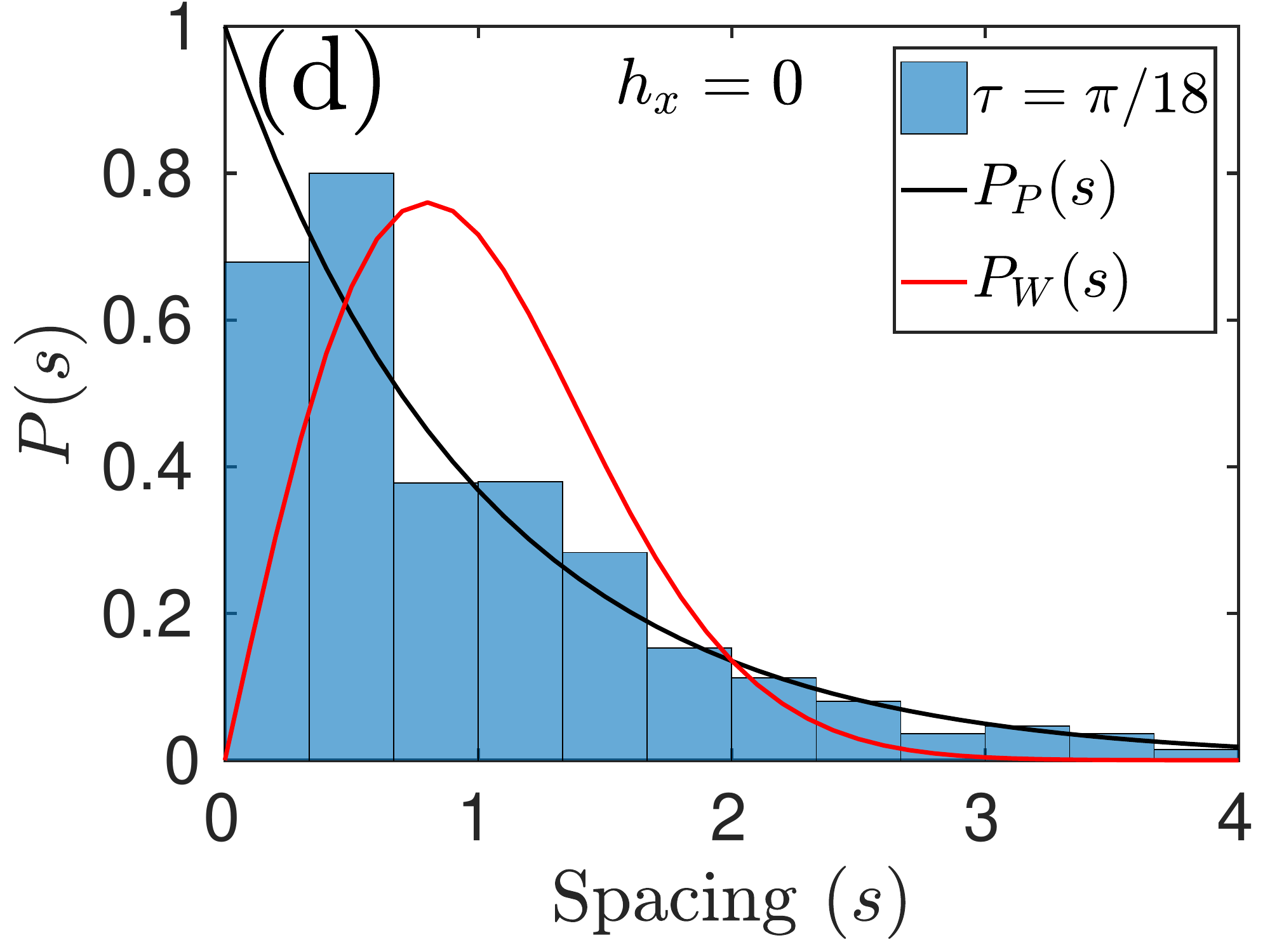}
       \caption{ Integrable $\mathcal{\hat U}_0$ system with parameters: $\tau=\pi/18$, $J_x=1$, $h_{x}=0$ and $h_{z}=4$.  \textbf{(a)} $C(n)/C(\infty)$ generated by SBOs vs. $n$ for $N=18$ and $\tau=\pi/18$, $3\pi/18$.  ($\log-\log$). Line with points represents data from the numerical calculation and solid line is the polynomial fitting. Inset shows  $1-C(n)/C(\infty)$ vs. $n$ ($\log-$linear). \textbf{(b)} $C(n)/C(\infty)$ vs. $n$ for $N=12$ and RBOs as observables.  \textbf{(c)} $1-C(n)/C(\infty)$ vs. $n$ for $N=12$ and RBOs as observables. Line with points is data generated numerically and solid line is the exponential fitting. \textbf{(d)} NNSD for $N=12$. In all the case open boundary condition is considered.}
       \label{pi18_hx0_hz4_int_N_p0} 
\end{figure*}


\begin{figure*}[hbt!] 
\includegraphics[width=.30\linewidth, height=.25\linewidth]{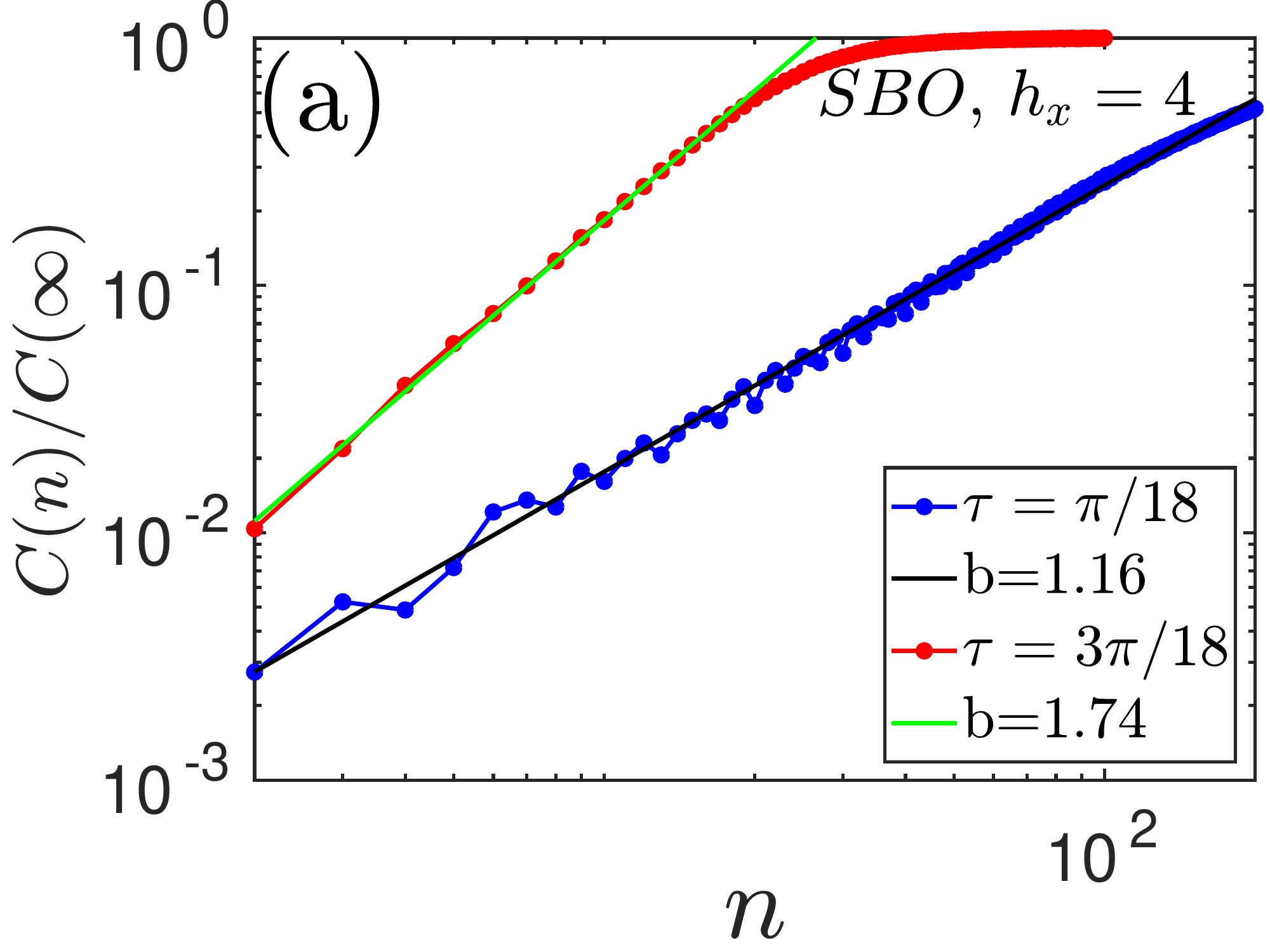}
\includegraphics[width=.30\linewidth, height=.25\linewidth]{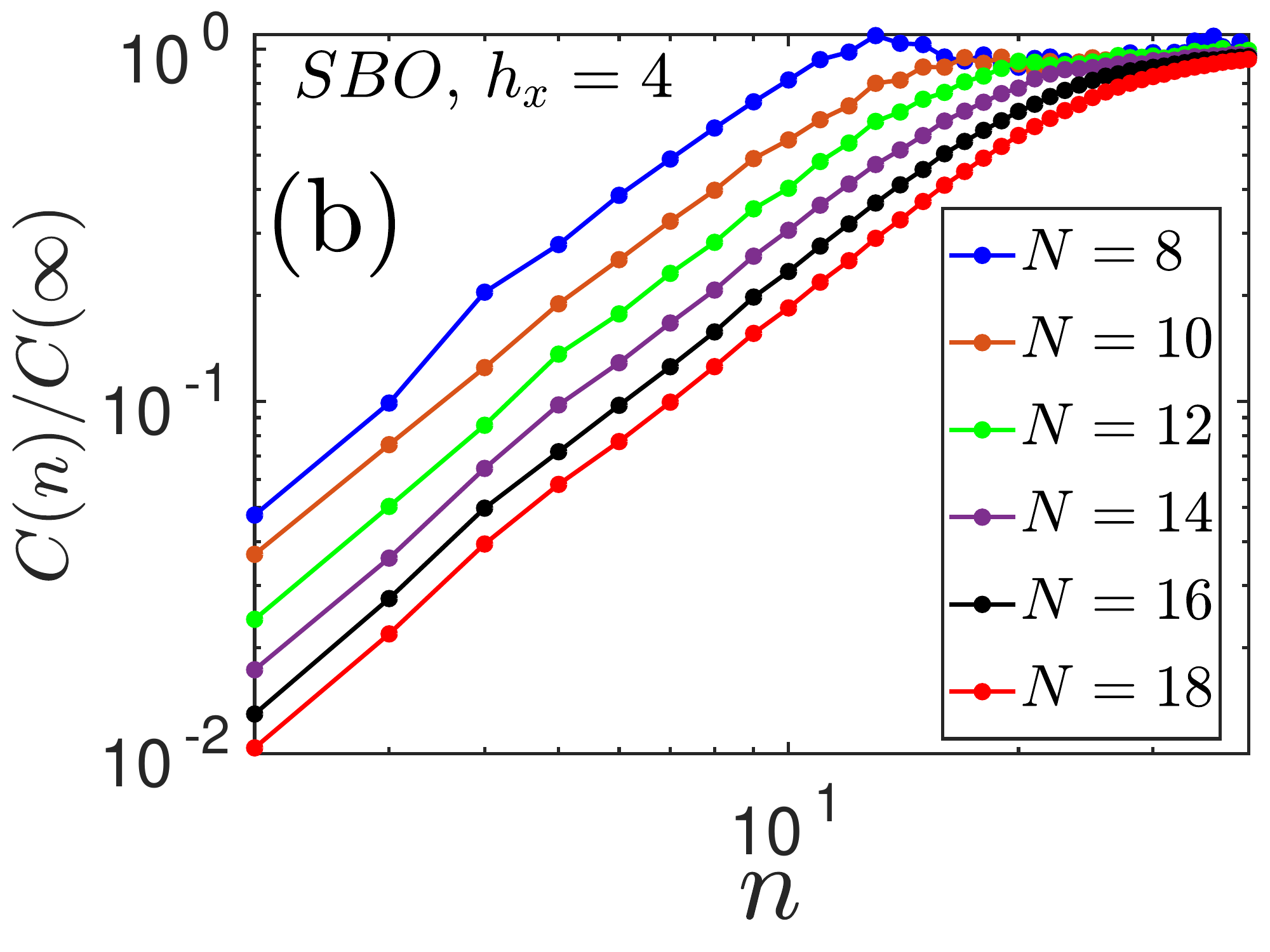}
\includegraphics[width=.30\linewidth,height=.25\linewidth]{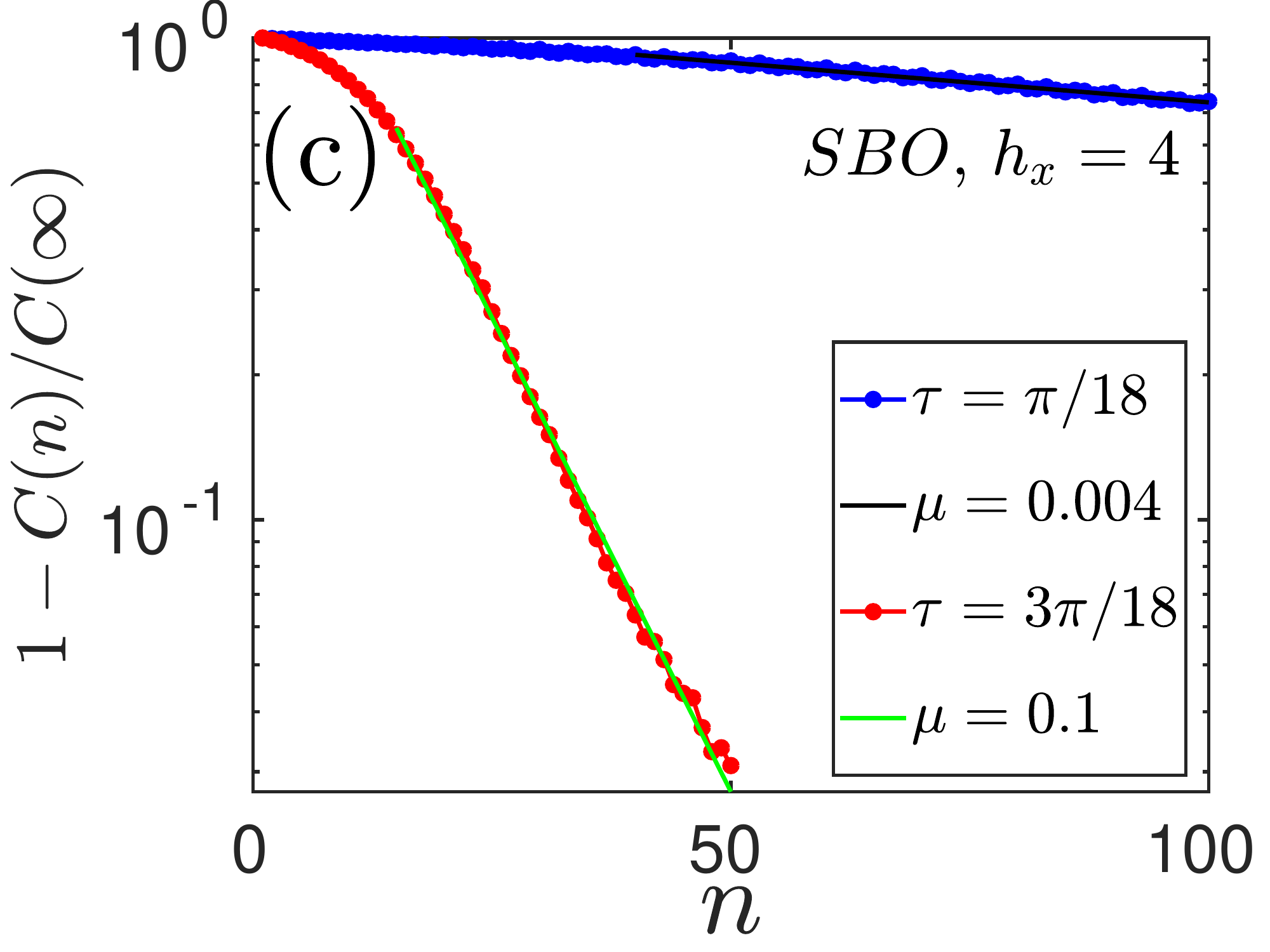}
\includegraphics[width=.45\linewidth, height=.30\linewidth]{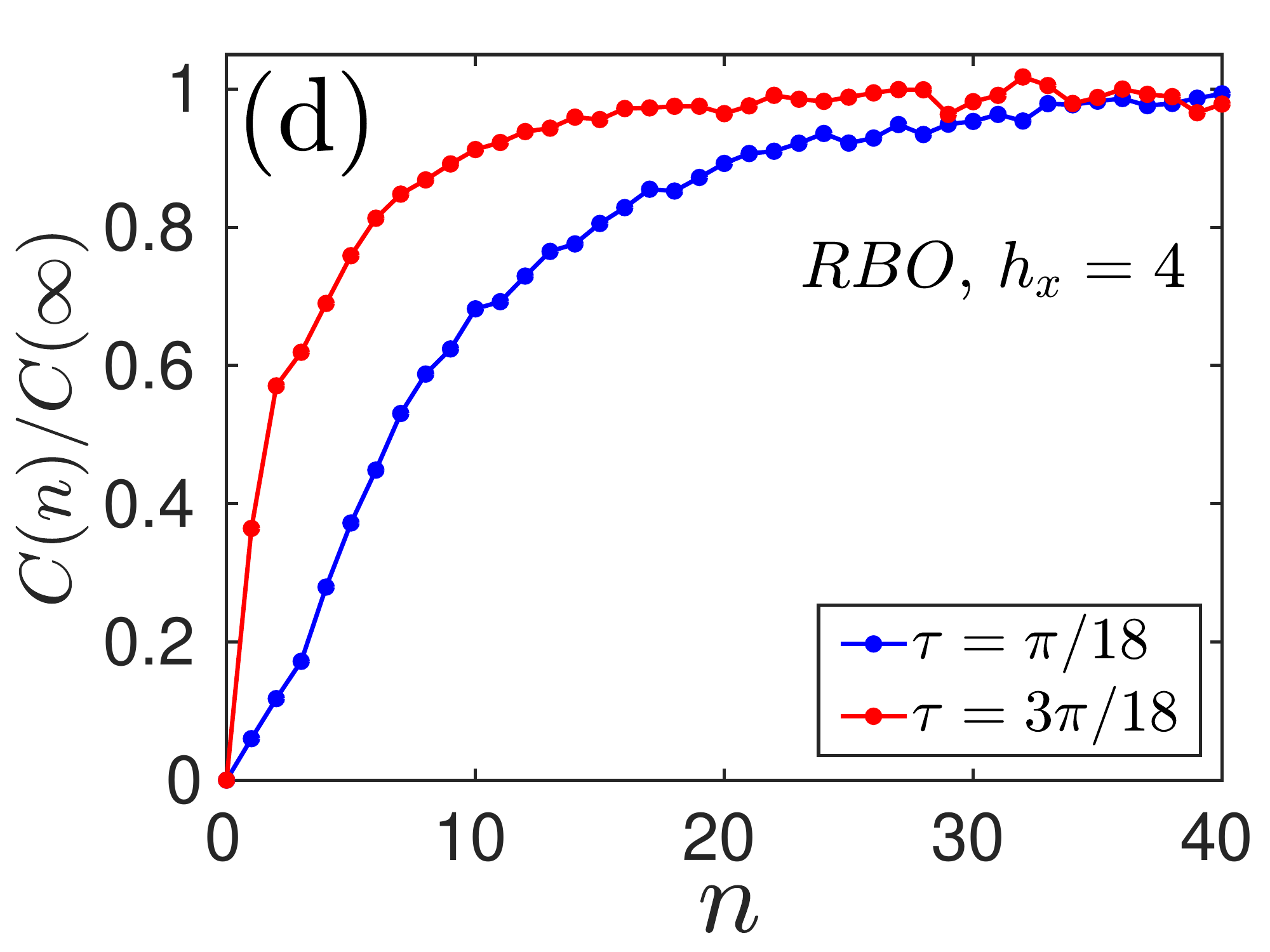}
\includegraphics[width=.45\linewidth, height=.30\linewidth]{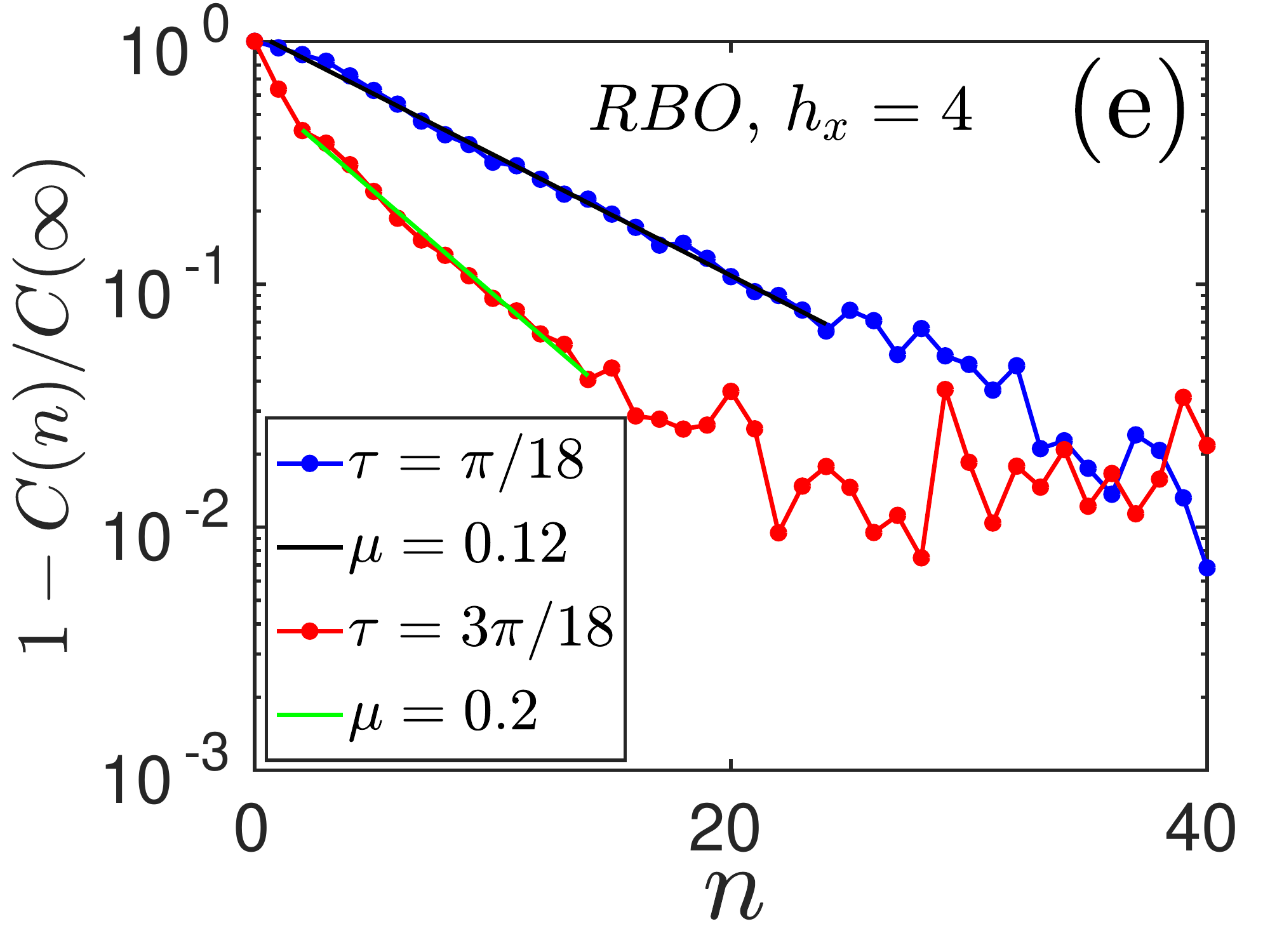}
\includegraphics[width=.45\linewidth,height=.30\linewidth]{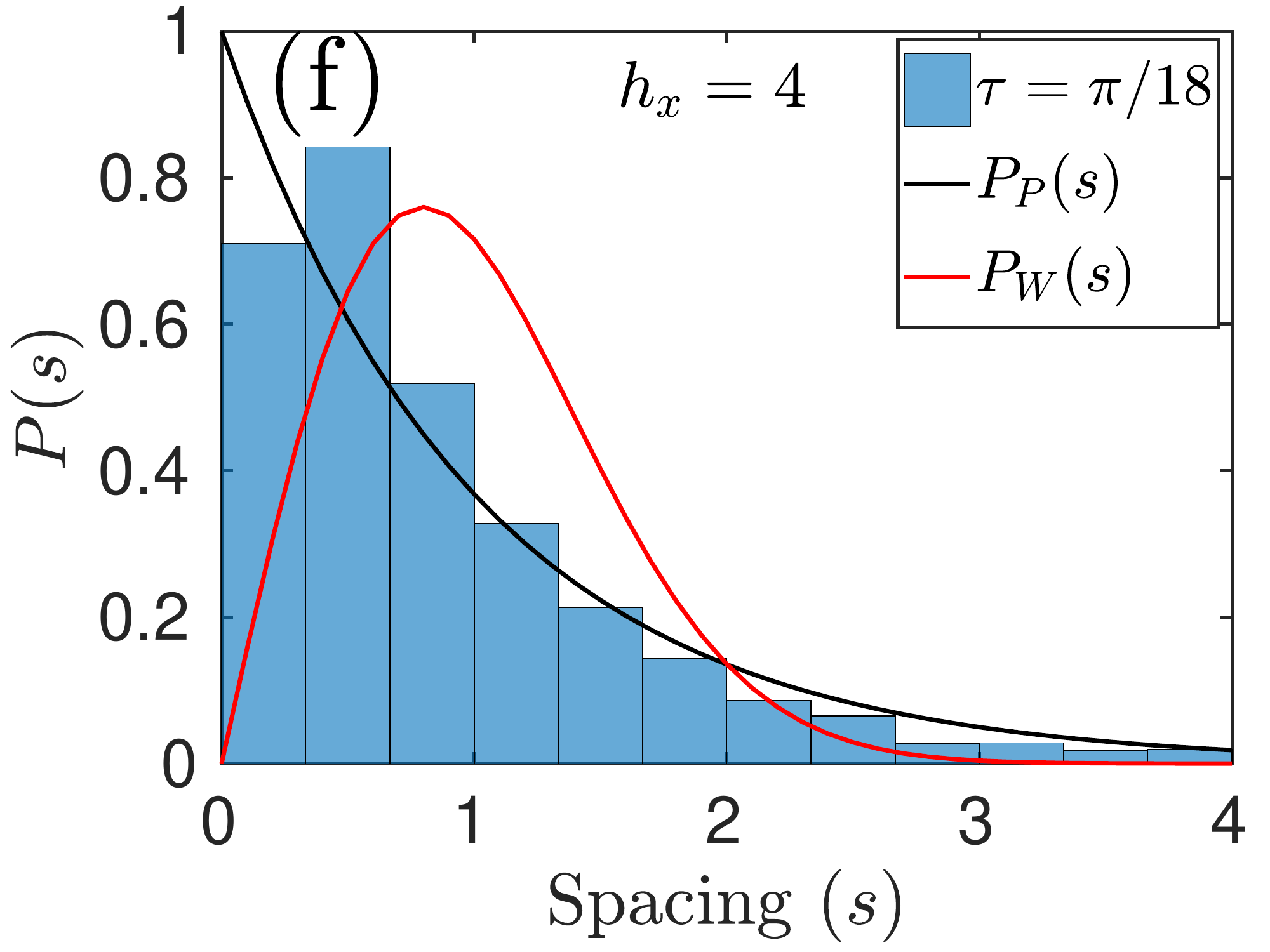}
\includegraphics[width=.45\linewidth,height=.30\linewidth]{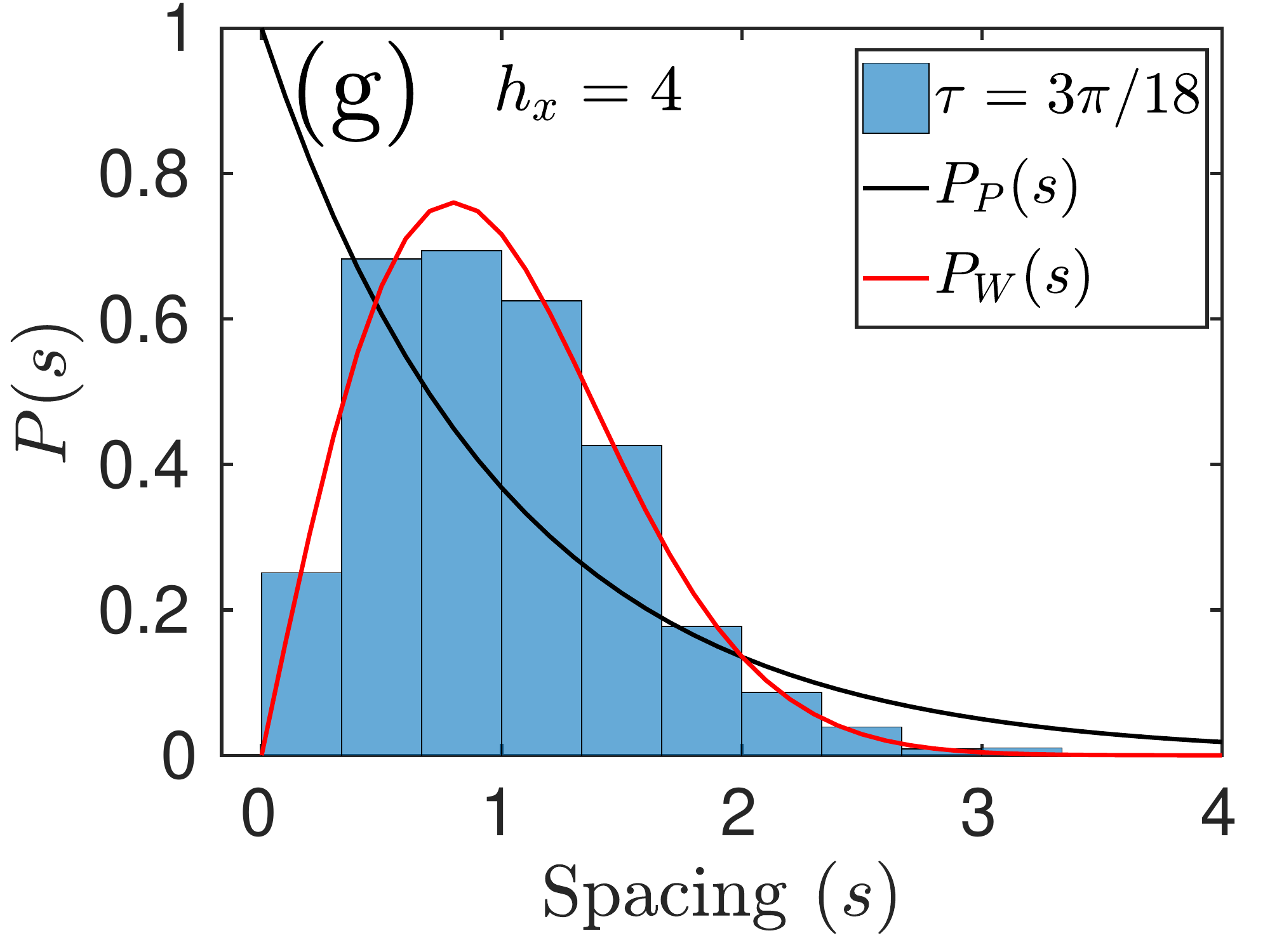}
\caption{Nonitegrable $\mathcal{\hat U}_x$ system with parameters:  $J_x=1$, $h_{x}=4$, $h_{z}=4$ and $\tau=\pi/18$, $3\pi/18$.  \textbf{(a)} Illustrates the $C(n)/C(\infty)$ by using the SBOs  vs. $n$ for $N=18$ ($\log-\log$). Lines with points represent data from the numerical calculation and solid lines are the polynomial fitting with exponent $b\approx1.18$ at $\tau=\pi/18$  and $b\approx1.74$ at $\tau=3\pi/18$. \textbf{(b)} $C(n)/C(\infty)$ by using the SBOs  vs. $n$ at different $N$ for $\tau=3\pi/18$. \textbf{(c)} $1-C(n)/C(\infty)$ vs. $n$ ($\log-$linear). Lines with points are data generated numerically and solid lines are the exponential fitting.   \textbf{(d)} Illustrates the OTOCs of RBOs vs. $n$ for $N=12$ \textbf{(g)} $1-C(n)/C(\infty)$ vs. $n$ ($\log-$linear). Lines with points are data generated numerically and solid lines are the exponential fitting. NNSD of the $\mathcal{\hat U}_x$ system at  \textbf{(f)} $\tau=\pi/18$ and \textbf{(g)} $\tau=3\pi/18$ with $N=12$. In all the cases open boundary chain is considered.}
\label{pi18_hx4_hz4_cf_nint_p0_N} 
\end{figure*}

\begin{figure*}[hbt!] 
\centering 
\includegraphics[width=.45\linewidth,height=.28\linewidth]{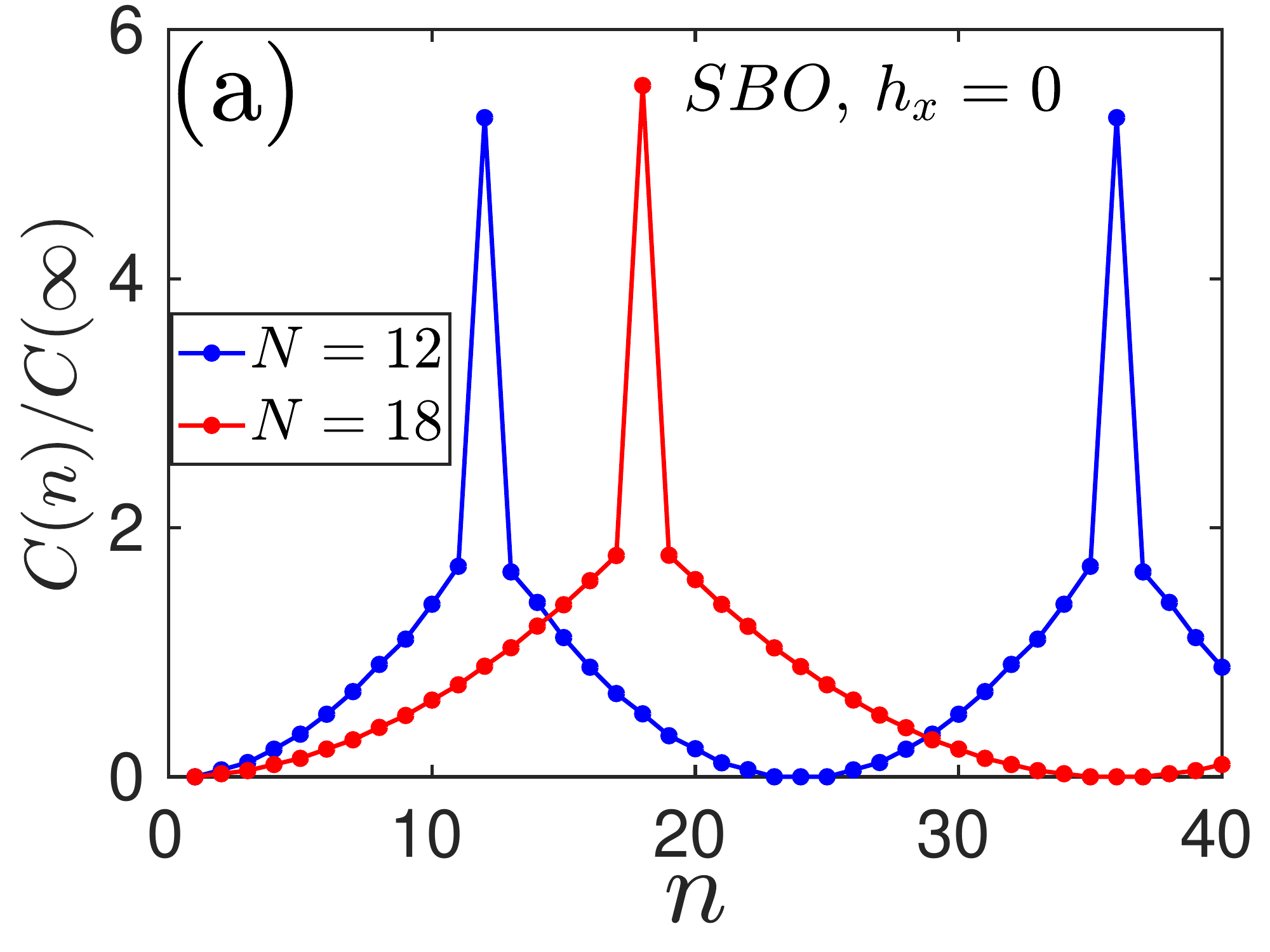}
\includegraphics[width=.45\linewidth,height=.28\linewidth]{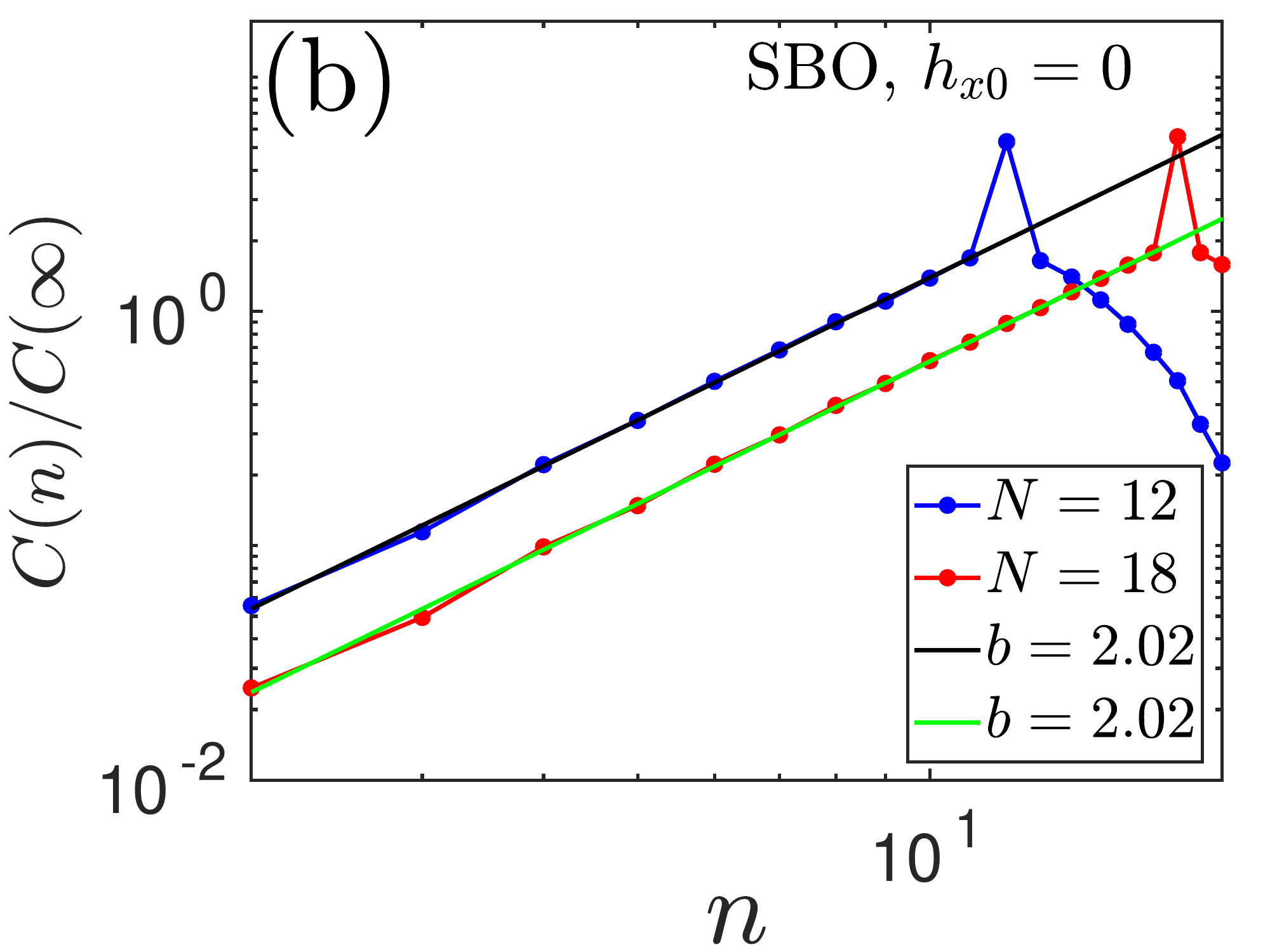}
\includegraphics[width=.45\linewidth, height=.28\linewidth]{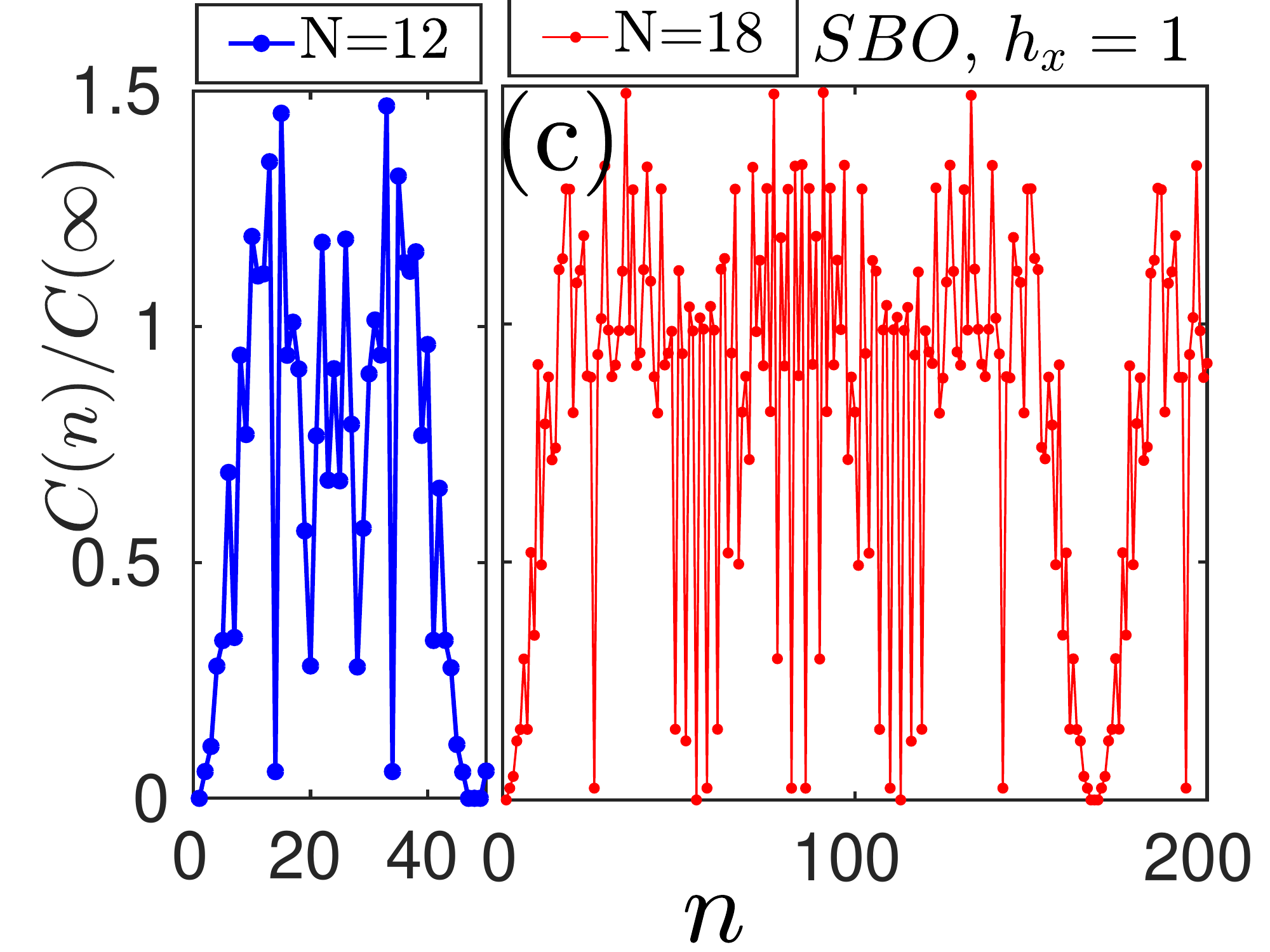}
\includegraphics[width=.45\linewidth, height=.28\linewidth]{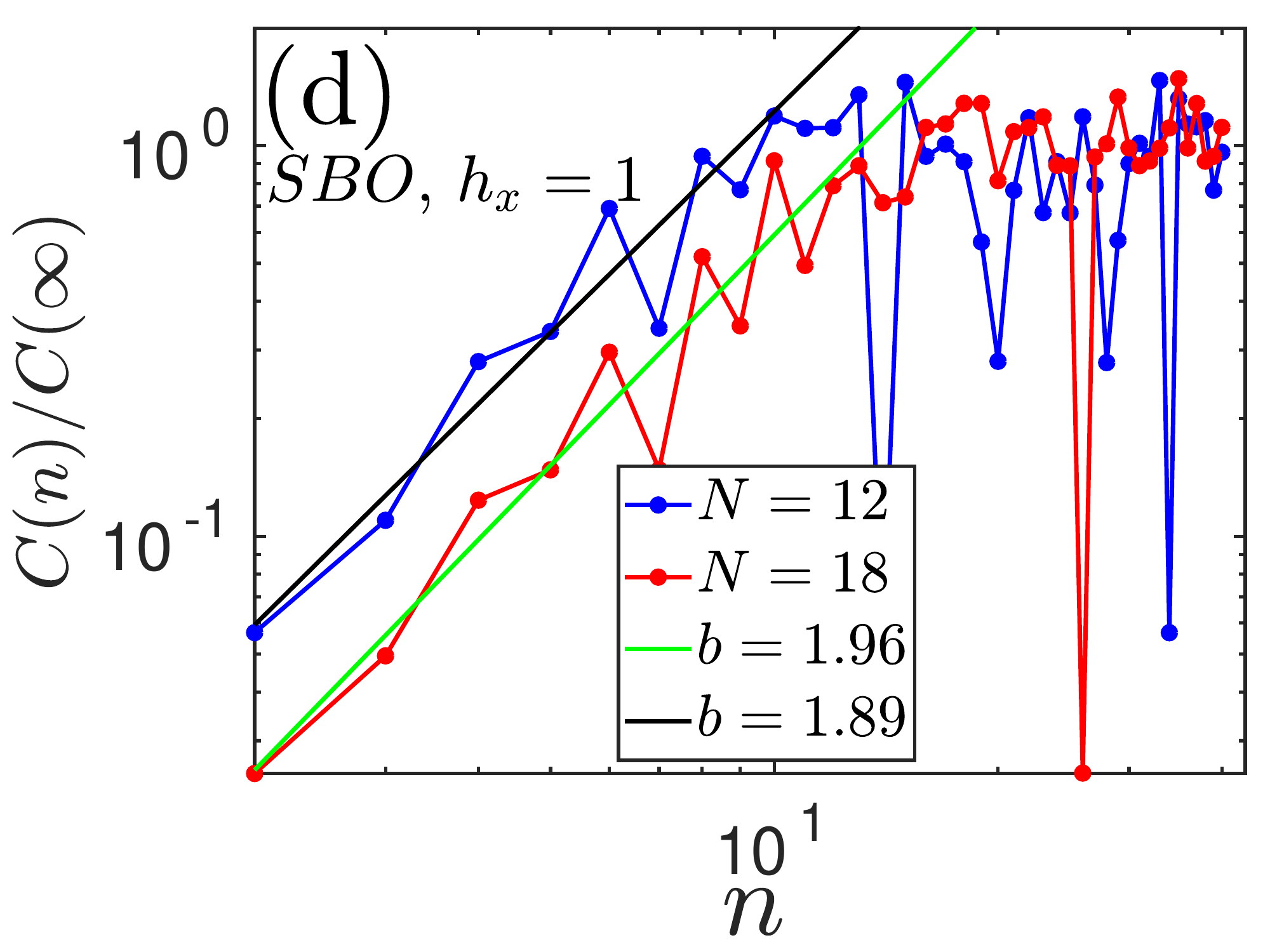}
\includegraphics[width=.45\linewidth, height=.28\linewidth]{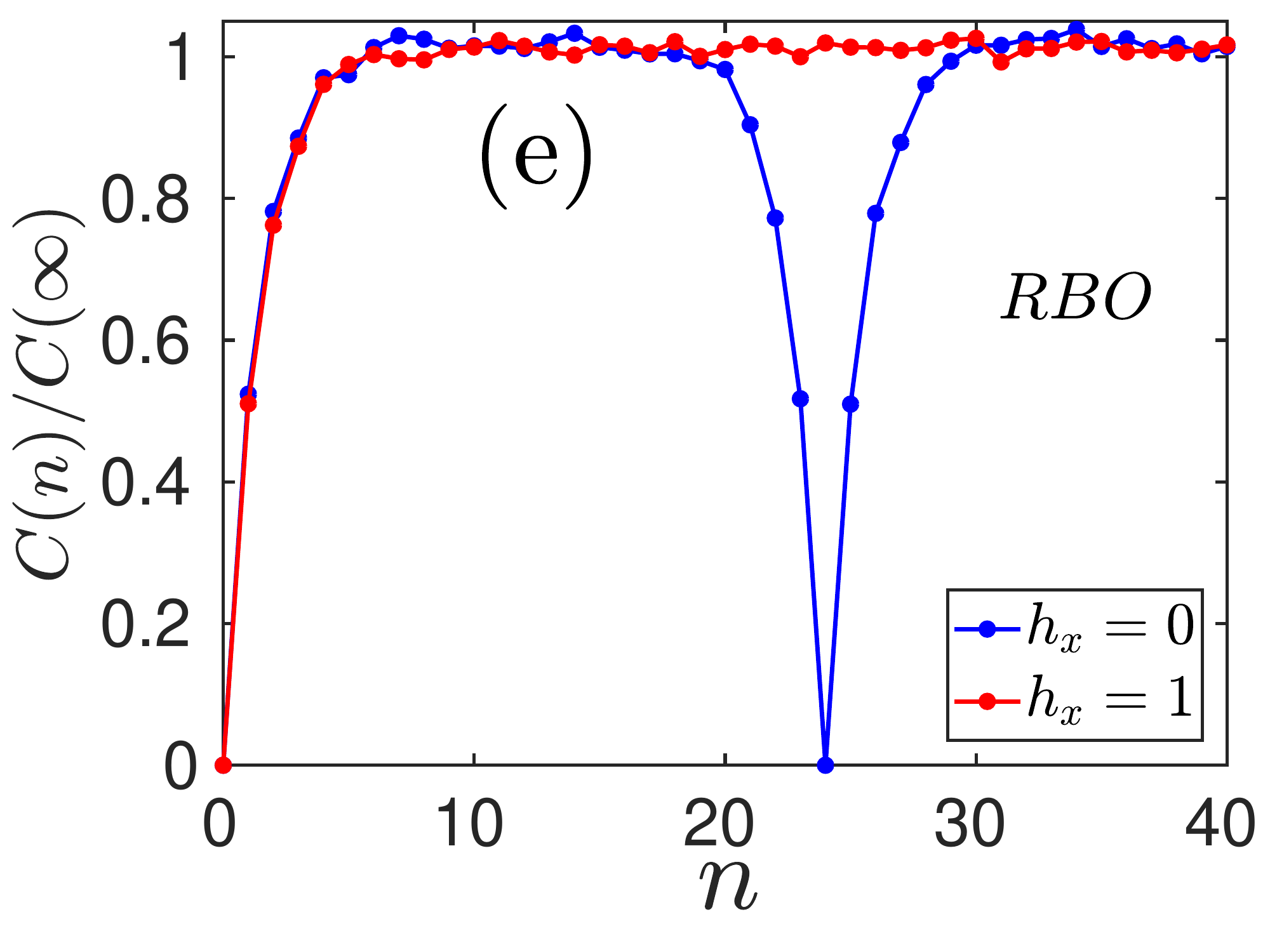}
\includegraphics[width=.45\linewidth, height=.28\linewidth]{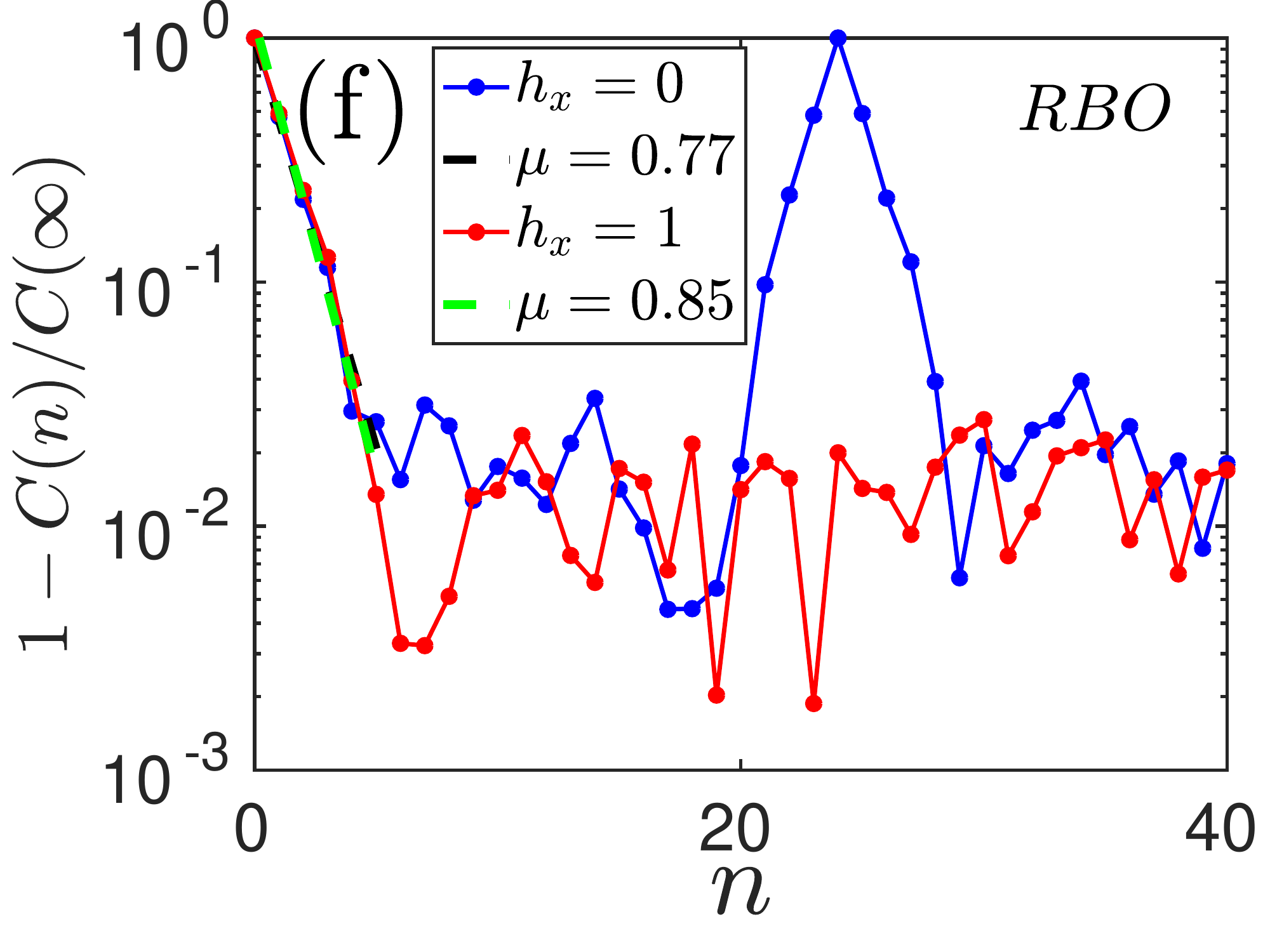}
\label{pi4_hx1_hz1_cf_log_rand5_N_nint_p0} 
\caption{ \textbf{(a)} $C(n)/C(\infty)$ of SBOs vs. $n$ in the $\mathcal{\hat U}_0$ system for $N=18$. \textbf{(b)} $\log-\log$ behaviour of ``\textbf{(a)}" in which lines with points represent data from the numerical calculation and solid lines are the polynomial fitting. \textbf{(c)} $C(n)/C(\infty)$ of SBOs with $n$ in the $\mathcal{\hat U}_x$ system for $N=18$. \textbf{(d)} $\log-\log$ behaviour of ``\textbf{(c)}" in which lines with points represent data from the numerical calculation and solid lines are the polynomial fitting. \textbf{(e)} $C(n)/C(\infty)$ of RBOs vs. $n$ in the $\mathcal{\hat U}_0$ and $\mathcal{\hat U}_x$ system for $N=12$. \textbf{(f)} $1-C(n)/C(\infty)$ vs. $n$ for $N=12$ ($\log-$linear). Lines with points are data generated numerically and solid lines are the exponential fitting. Other parameters: $J_x=1$, $h_{x}=0/1$, $h_{z}=1$ and  $\tau=\pi/4$. In all the case open boundary chain is considered.}
  \label{pi4_hx1_hz1_cf_nint_p0_N} 
\end{figure*}
\begin{figure*}[hbt!] 
\centering
\includegraphics[width=.30\linewidth,height=.22\linewidth]{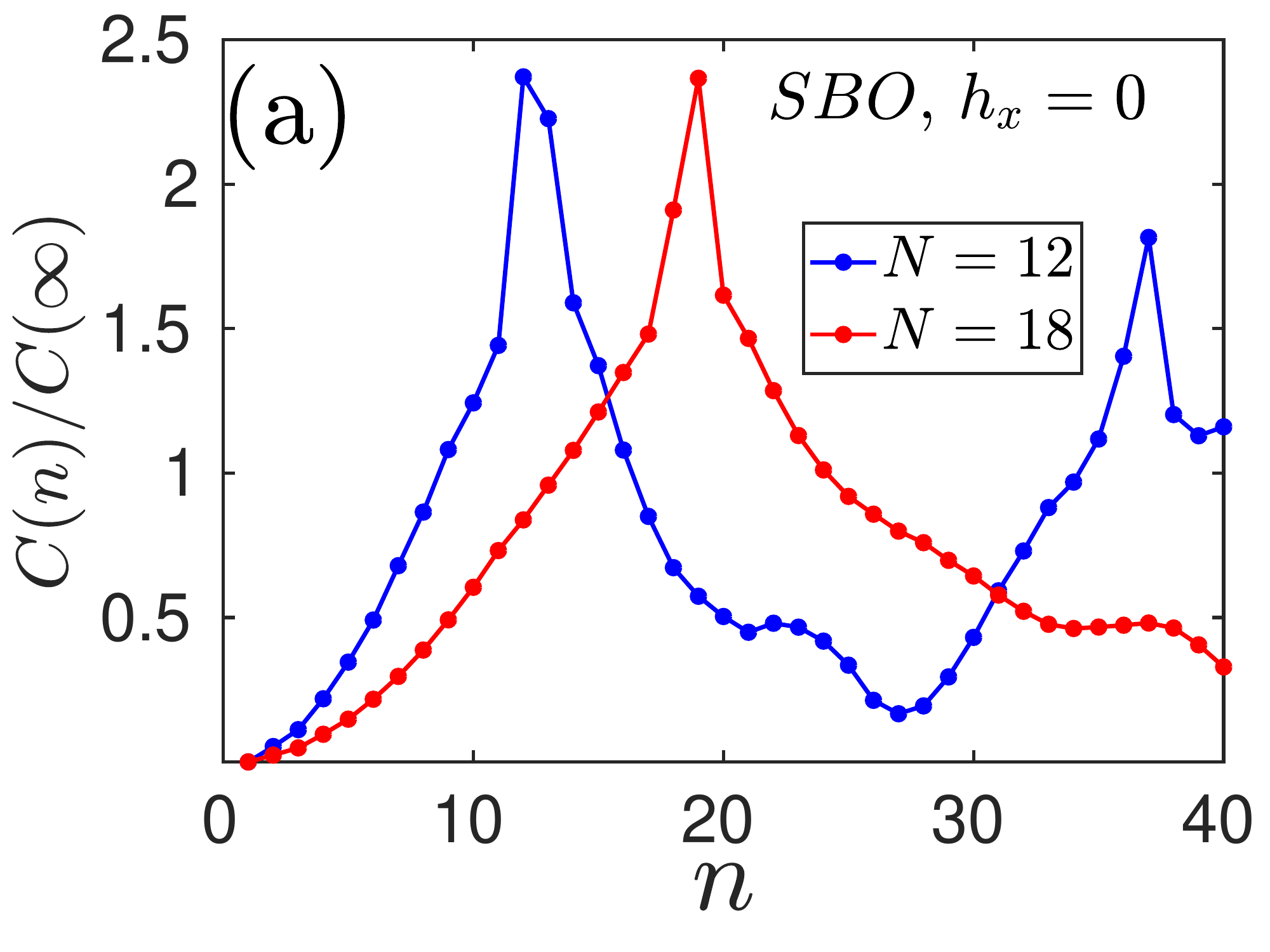}
  \includegraphics[width=.30\linewidth,height=.22\linewidth]{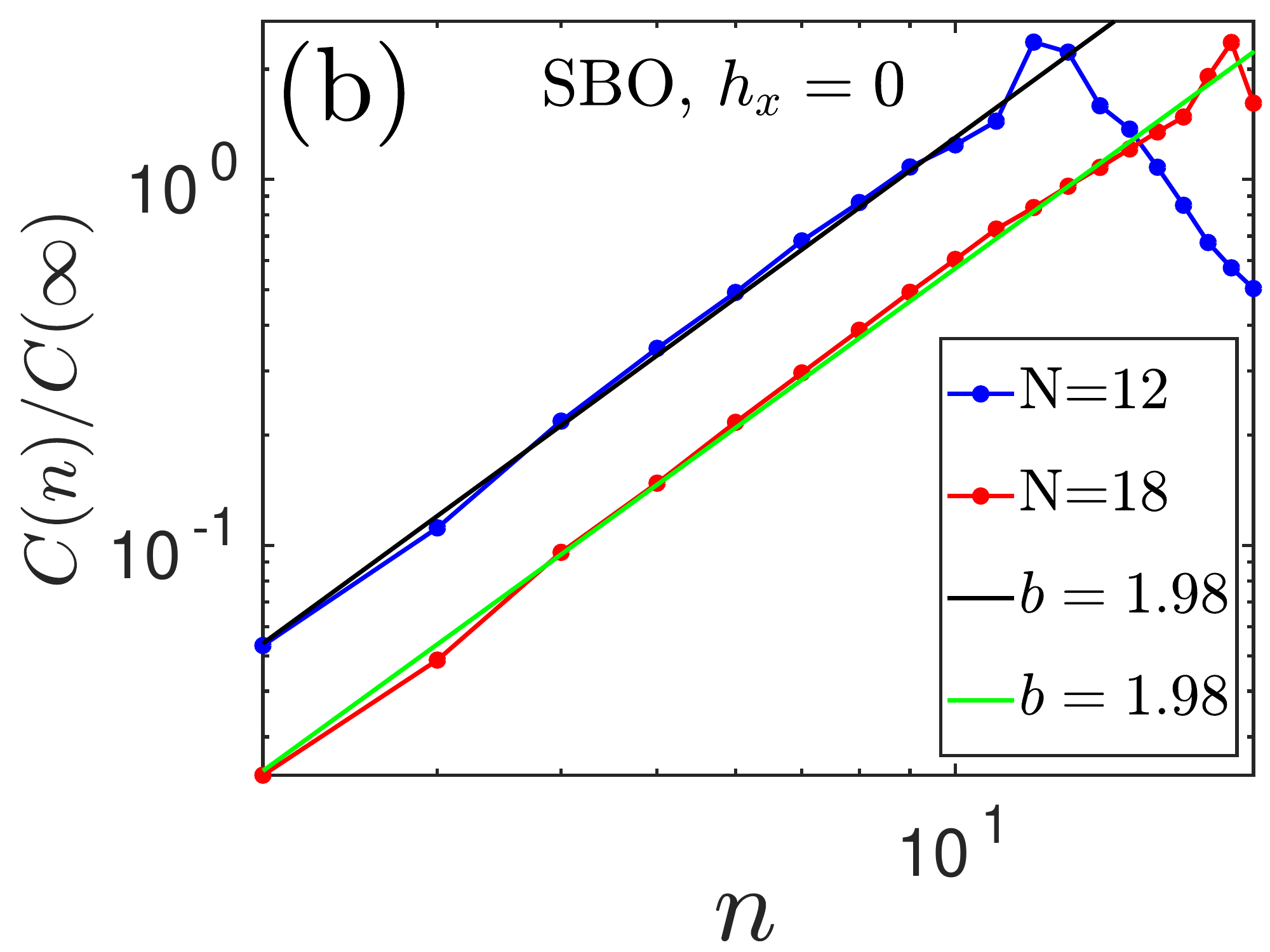}
  \includegraphics[width=.30\linewidth,height=.22\linewidth]{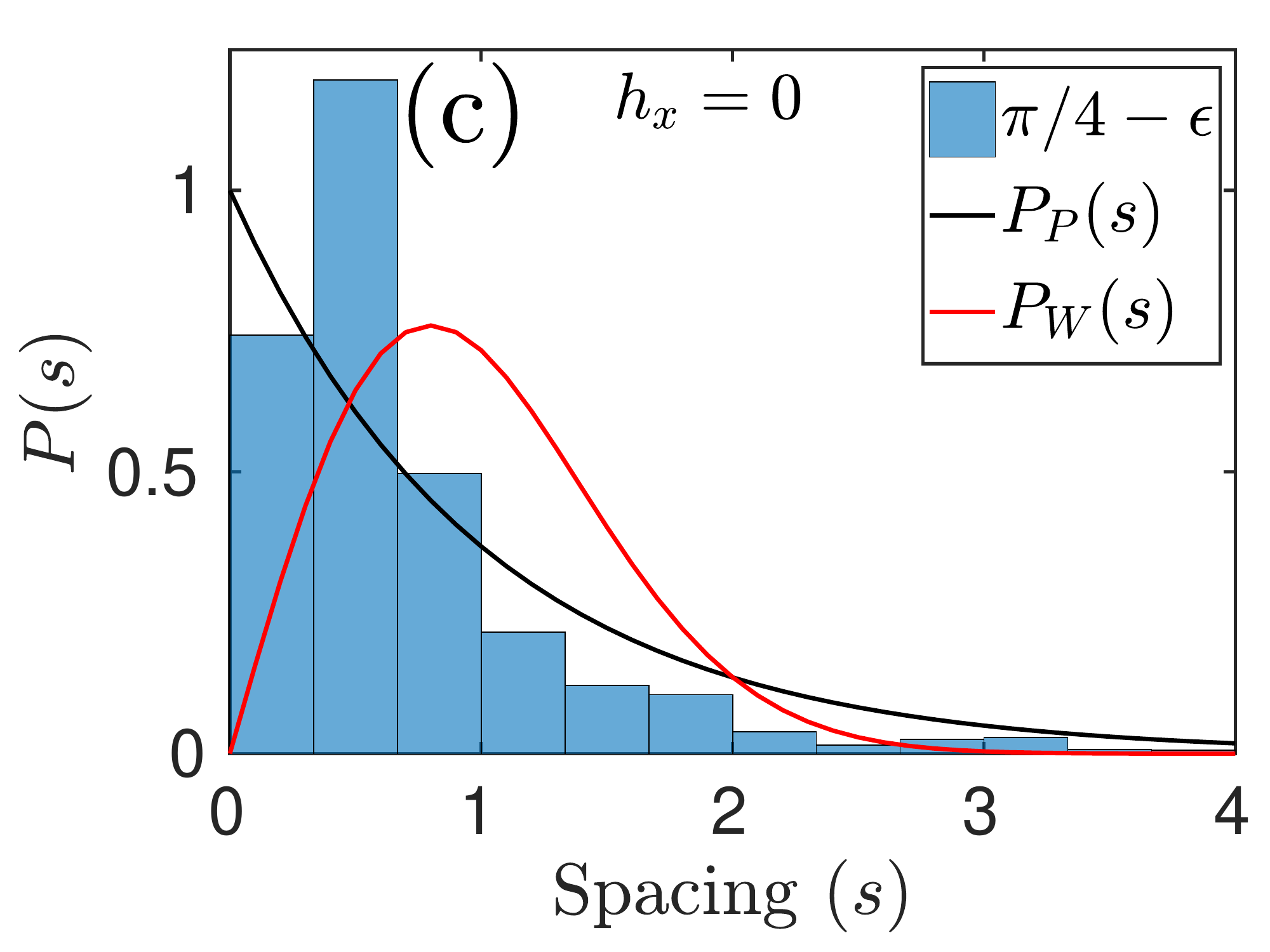}
\caption{Integrable $\mathcal{\hat U}_0$ system with parameters: $\tau=\pi/4-\epsilon(=\pi/50)$, $J_x=1$, $h_{x}=0$ and $h_{z}=1$.   \textbf{(a)} $C(n)/C(\infty)$ of SBOs vs. $n$ in the $\mathcal{\hat U}_0$ system for $N=18$. \textbf{(b)} $\log-\log$ behaviour of ``\textbf{(a)}" in which lines with points represent data from the numerical calculation and solid lines are the polynomial fitting. (\textbf{c}) NNSD of the integrable $\mathcal{\hat U}_0$ system with $N=12$.}
\label{pi4e_hx0_hz1_cf_int_p0_N} 
\end{figure*}
\begin{figure*}
\centering
\includegraphics[width=.30\linewidth,height=.25\linewidth]{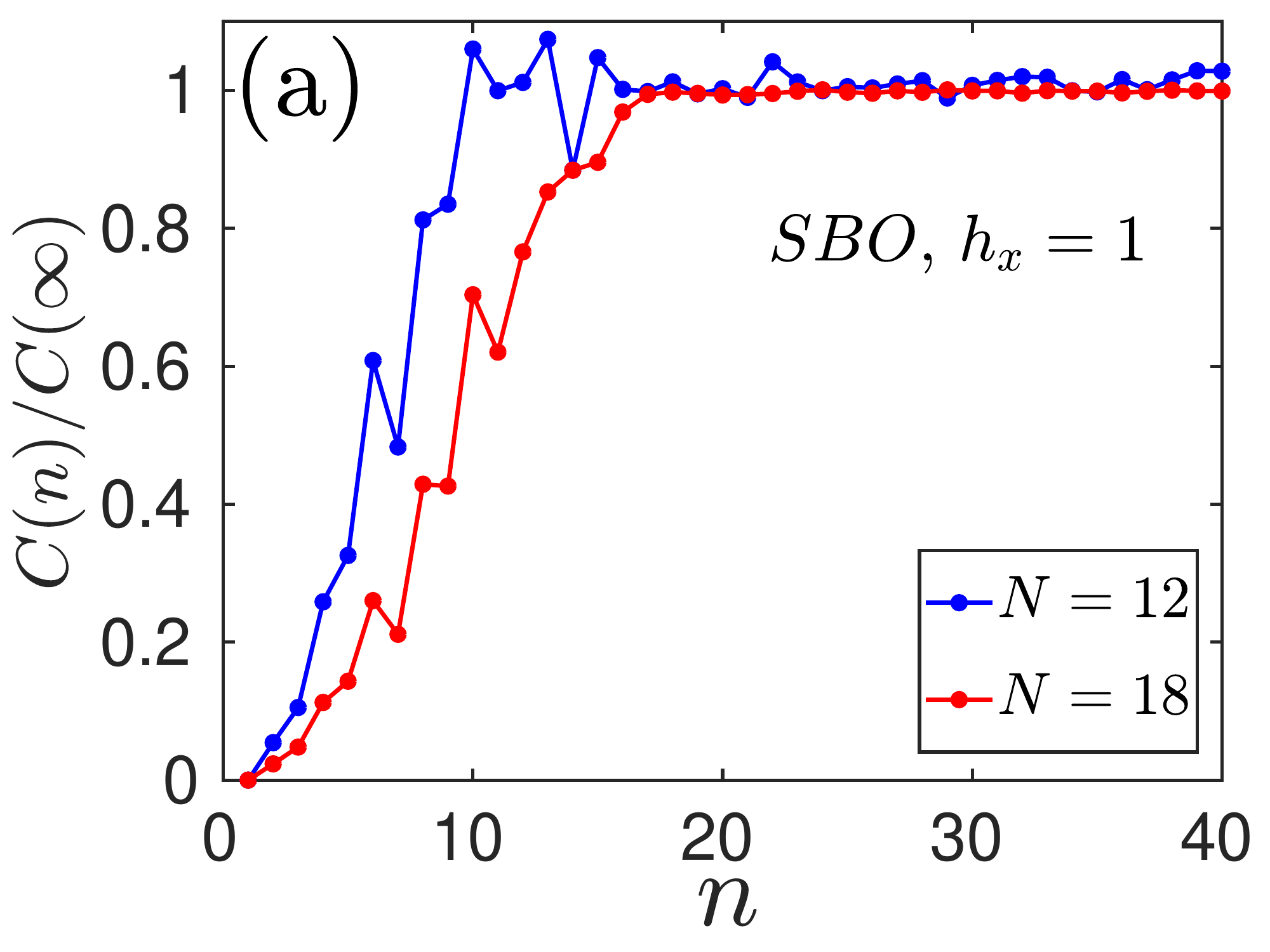}
  \includegraphics[width=.30\linewidth,height=.25\linewidth]{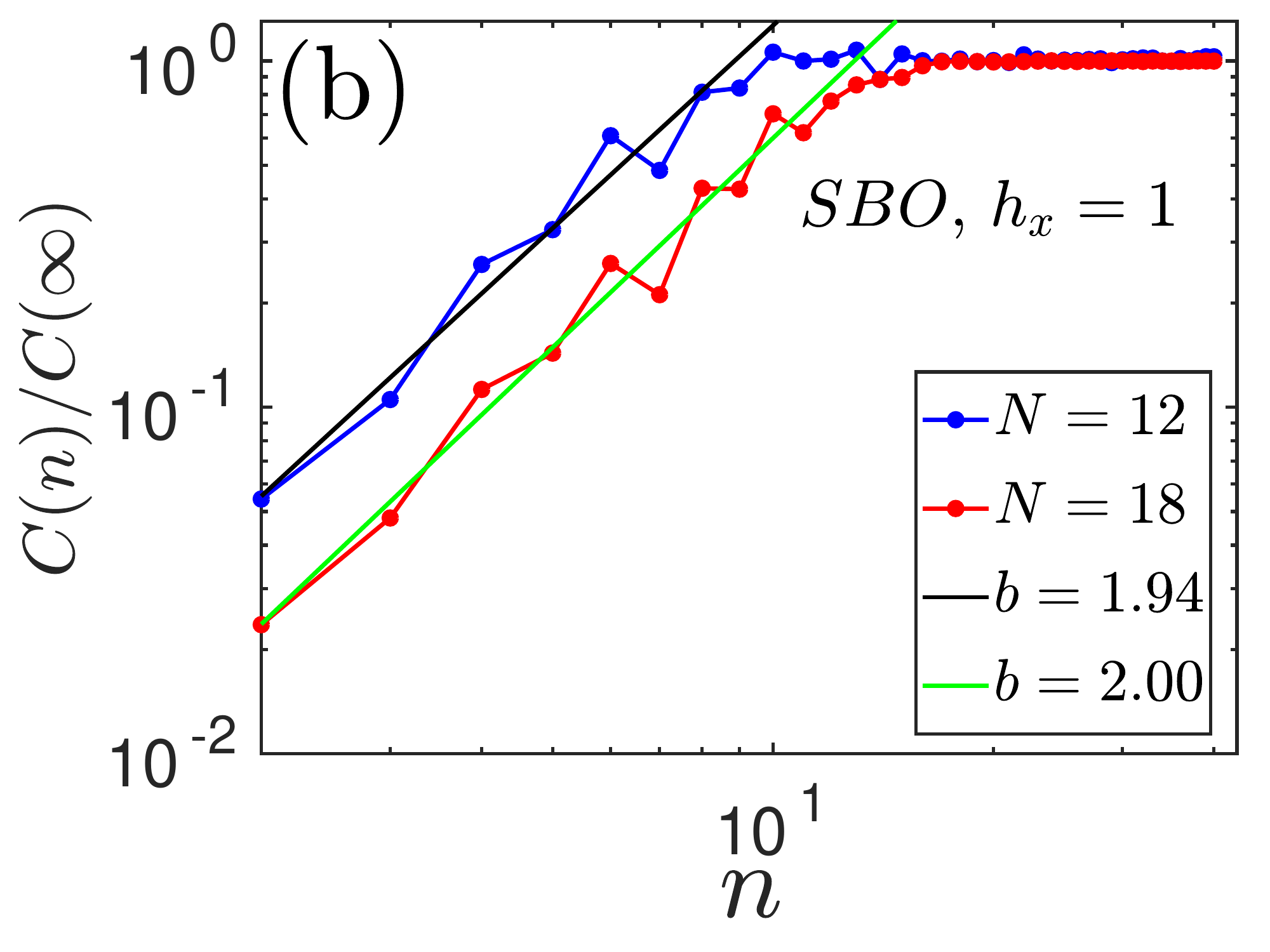}
  \includegraphics[width=.30\linewidth, height=.25\linewidth]{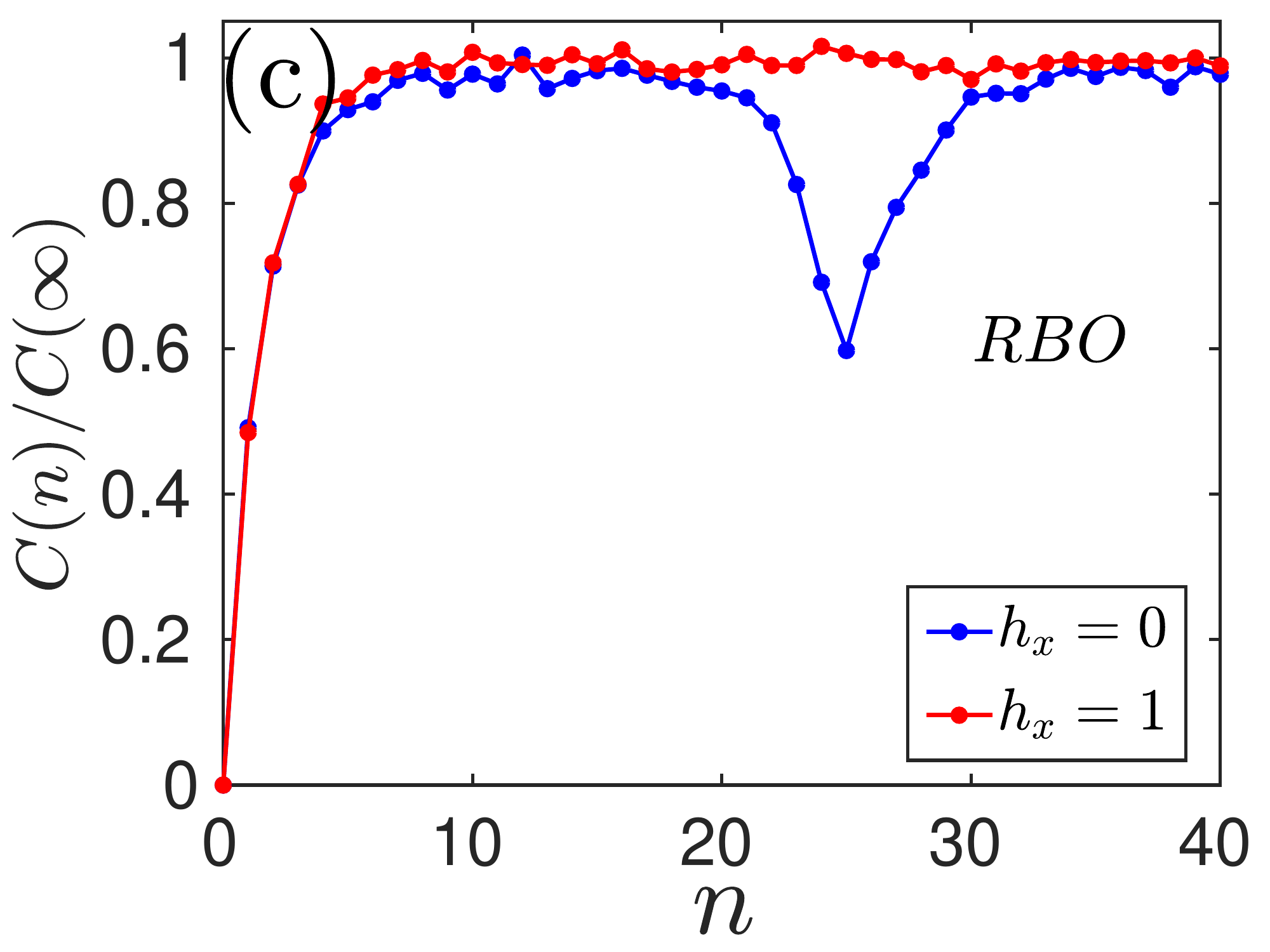}
\includegraphics[width=.45\linewidth, height=.30\linewidth]{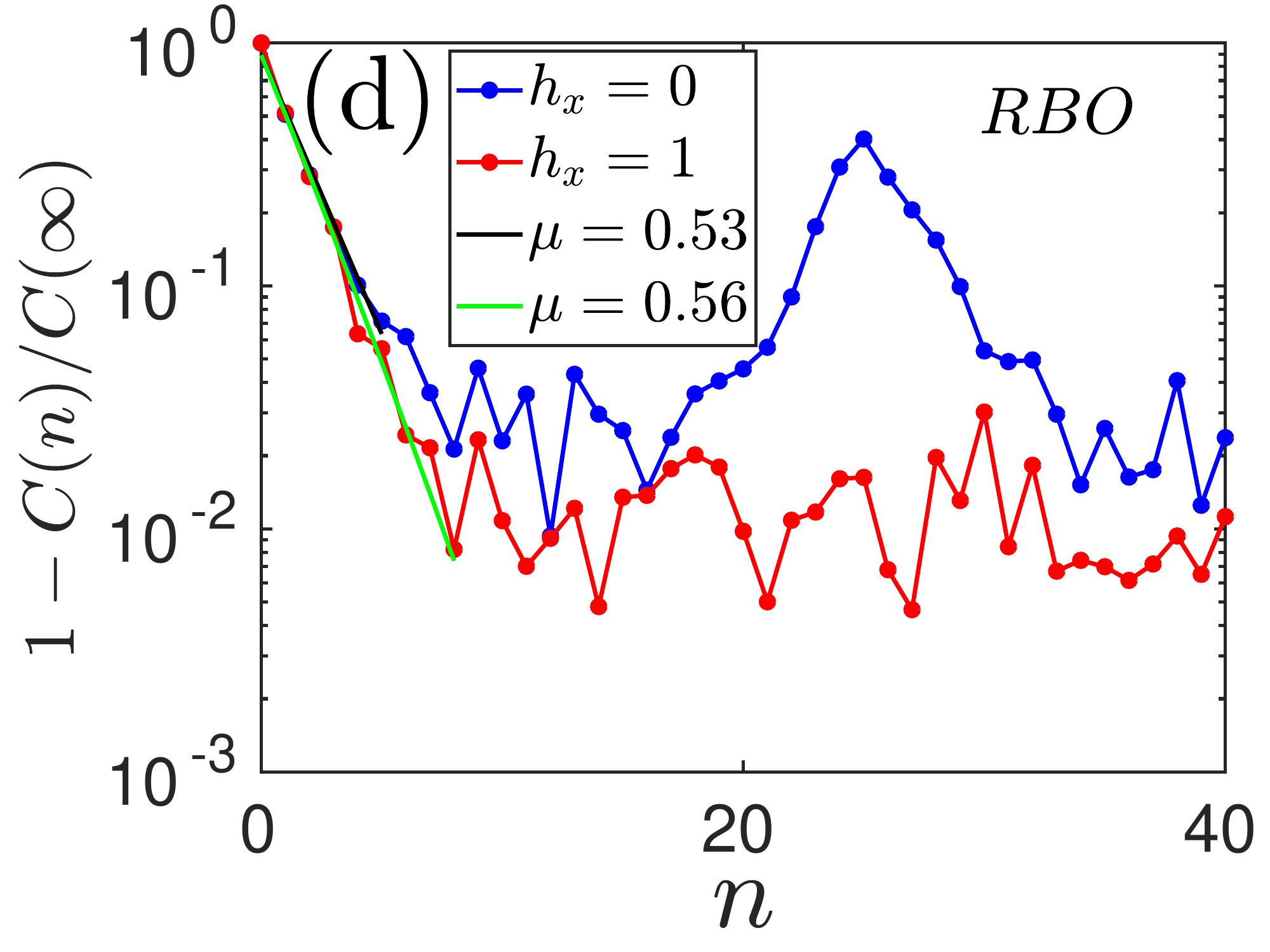}
\includegraphics[width=.45\linewidth,height=.30\linewidth]{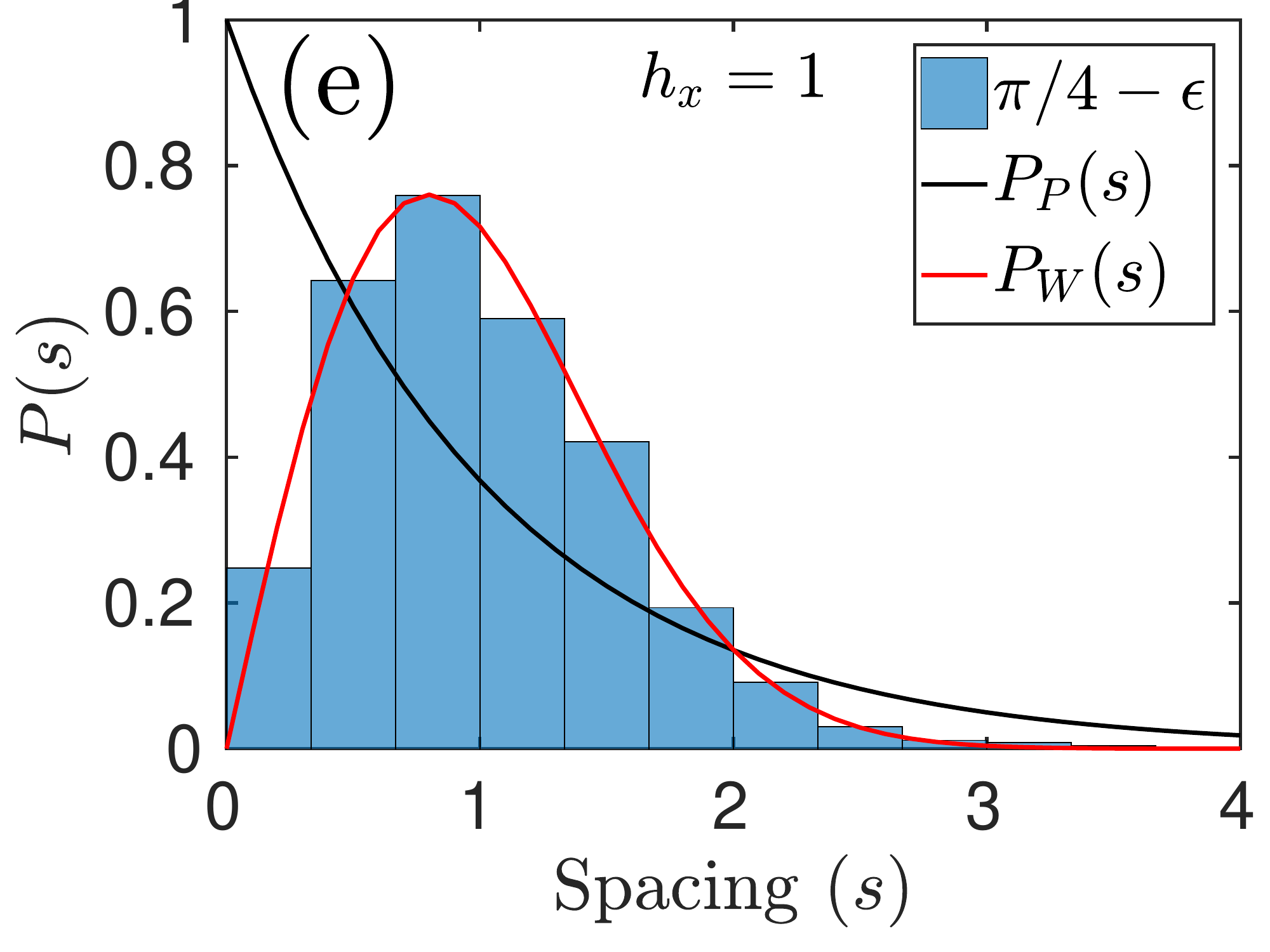}
 \caption{\textbf{(a)} $C(n)/C(\infty)$ of SBOs vs. $n$ in the $\mathcal{\hat U}_x$ system for $N=12$ and $18$. \textbf{(b)} $\log-\log$ behaviour of ``(\textbf{a})" in which lines with points represent data from the numerical calculation and solid lines are the polynomial fitting. \textbf{(c)} $C(n)/C(\infty)$ of RBOs vs. $n$ in the $\mathcal{\hat U}_0$ and $\mathcal{\hat U}_x$ system for $N=12$. \textbf{(d)} $1-C(n)/C(\infty)$ vs. $n$ for $N=12$ ($\log-$linear). Lines with points are data generated numerically and solid lines are the exponential fitting. \textbf{(d)} NNSD of the $\mathcal{\hat U}_x$ system for $N=12$. Other parameters: $J_x=1$, $h_{x}=0/1$, $h_{z}=1$ and  $\tau=\pi/4-\epsilon(=\pi/50)$.}
\label{pi4e_hx1_hz1_cf_nint_p0_N} 
\end{figure*}
\section{Constant field Floquet system}
\label{CF}
 We analyze the OTOC given by Eq.~(\ref{Cn1}) for the integrable $\mathcal{\hat{U}}_0$ and the nonintegrable $\mathcal{\hat{U}}_x$ systems defined in section \ref{model}. The value of the magnetic fields is fixed at $h_x=0,\, h_z=4$ for the integrable case and $h_x=4,\, h_z=4$ for the non-integrable case. This the Floquet period $\tau$ acts as a parameter to drive the system into different dynamical regimes. In this manuscript, we will discuss the dynamic (pre-scrambling time) and saturation (post-scrambling time) regions of OTOC, generated by spin-block operators defined in Eq. (\ref{Block}) as well as random operators referred to as RBO for ``random block-operators". 

In the integrable case $\mathcal{\hat{U}}_0$ the dynamic region of the OTOC shows power-law growth, $C(n)/C(\infty)\sim n^{b}$, with the exponent being $b\approx 2$. This is shown in  [Fig. \ref{pi18_hx0_hz4_int_N_p0}(a)] for two values of the period, $\tau=\pi/18$, and $3\pi/18$. For period $0<\tau<\pi/2$, the OTOC shows power-law growth with the same approximate quadratic growth, except at $\tau=\pi/4$ at which it vanishes. However the OTOC does not saturate at any particular value beyond the scrambling time as can be seen in inset of  Fig. \ref{pi18_hx0_hz4_int_N_p0}(a). 
\par

Replacing the spin-operators with random block observables, the OTOC thermalizes quickly as compared to SBOs. This leads to disappearance of the power-law growth for $\tau=\pi/18$ [Fig. \ref{pi18_hx0_hz4_int_N_p0}(b)], and is replaced by an exponential saturation $C(n)/C(\infty)\sim 1-\exp(-\mu n)$, with the rate $\mu\approx0.14$ [Fig. \ref{pi18_hx0_hz4_int_N_p0}(c)]. The OTOC averaged over the random matrices $\hat V$ and $\hat W$ drawn from GUE for  $\mathcal{\hat{U}}_0$  system is exactly same as OPEE $E_l[\mathcal{\hat{U}}_0]$, as established in Eq. (\ref{OPEE_OTOC}).
\par
Fig.~\ref{pi18_hx0_hz4_int_N_p0}(d) shows that the NNSD of the integrable $\mathcal{\hat U}_0$ system at 
$\tau=\pi/18$  is Poisson type rather than Wigner-Dyson type \cite{Fortes2019,lin2018out}. 
The system displays Poisson statistics at all the Floquet periods from $0<\tau<\pi/2$ except at $\pi/4$. 
At $\tau=\pi/4$, as $h_z=4$, the field term is effectively absent and  
$\mathcal{\hat U}_x =e^{-i\hat H_{xx}\frac{\pi}{4}}$, leading to vanishing OTOC,
for the choice of spin observables.

\par
OTOC in the nonintegrable $\mathcal{\hat U}_x$ system shows a power-law growth before the scrambling time, similar to that in the integrable $\mathcal{\hat U}_0$ case. However, in the nonintegrable case, the exponent of the power-law is smaller as compared to the integrable case and the exponent increases with increasing $\tau$. In order to extract the effects of nonitegrability we focus on two $\tau$ values: $\tau=\pi/18$ and $3\pi/18$.  At $\tau=\pi/18$ and $3\pi/18$ exponents of the power-law are $b\approx 1.18$ and $b\approx1.74$, respectively [Fig. \ref{pi18_hx4_hz4_cf_nint_p0_N}(a)]. Hence, at $\tau=3\pi/18$, the exponent is nearly quadratic in a power-law growth and independent of the system size, but the scrambling time of the OTOC depends on the system size. Larger the size, longer is the scrambling time. Hence, the scrambling time of OTOC exhibits the finite-size effect as shown in Fig. \ref{pi18_hx4_hz4_cf_nint_p0_N}(b). In a thermodynamic limit, we expect the scrambling time  to occur after infinite number of kicks. OTOC approaches to saturation exponentially at any $\tau$, however, the rate of saturation increases with increasing $\tau$ [ see Fig. \ref{pi18_hx4_hz4_cf_nint_p0_N}(c)].
\par
Now, if we  replace the localized spin observables $\hat V$  and $\hat W$ to pre-scrambled random block observables, the growth of OTOC does not show Lyapunov or power-law type at any $\tau$ [Fig. \ref{pi18_hx4_hz4_cf_nint_p0_N}(d)]. It is exactly same as   OPEE  $E_l[\mathcal{\hat{U}}_x]$, as given by Eq. (\ref{OPEE_OTOC}). OTOC saturates exponentially and the rate $\mu$ is $\approx 0.12$ for $\tau=\pi/18$ and $\approx0.20$ for $3\pi/18$ as shown in Fig.~\ref{pi18_hx4_hz4_cf_nint_p0_N}(e). This is correlated with quantum chaos being prevalent at $\tau=3 \pi/18$, while $\tau=\pi/18$ seems to be near-integrable.

\par 
This is consistent with the fact that NNSD of the nonintegrable Floquet system displays nearly Poissonian distribution at $\pi/18$ and Wigner-Dyson distribution at Floquet period $3\pi/18$ and moves towards Poisson distribution as the Floquet periods increases further from $3\pi/18$ to $\pi/4$. Therefore, we find $\tau=3\pi/18$ as the most chaotic point in the Floquet system [Fig.~\ref{pi18_hx4_hz4_cf_nint_p0_N} (f, g)] in terms of NNSD. 

\par
The Floquet Ising model is special at $J_x \tau= \pi/4$ which was reported in different contexts earlier. With the choice of appropriate magnetic fields such systems can show exact ballistic growth of block entanglement, revivals and so on \cite{Mishra2015,naik2019controlled,Pal2018}. We will study a nontrivial example of this in the next section, in the context of OTOCs.

\section{Special case: $h_{z}=1, h_{x}=0,1$, $\tau=\pi/4$}
\label{special_CF}
In the Ising Floquet system, there is a peculiar set of parameters {\it{viz.}} when $\tau=\pi/4$ for both the integrable $\mathcal{\hat U}_0$ case with $(h_x=0, h_z=1)$ and nonintegrable case $\mathcal{\hat U}_x$ with $(h_x=1,\, h_z=1)$. At this particular set of parameters, OTOC shows periodic oscillation in both integrable, as well as  nonintegrable systems. In the integrable case, OTOC oscillates with a time period equal to $2N$. 

It attains a maximum value at $n=(2m-1) N$ and goes to zero at $n=2 m N$, where $m\in \mathbb{Z}^+$ [Fig. \ref{pi4_hx1_hz1_cf_nint_p0_N}(a)]. The maximum value obtained is several times the saturation value of the nonintegrable case, namely $C(\infty)$.
OTOC shows quadratic growth ($\sim n^b$, $b\approx 2$) till $N-1$ kicks and the exponent is independent of the system size [Fig. \ref{pi4_hx1_hz1_cf_nint_p0_N}(b)]. 

It should be noted that both the entanglement entropy of quenches and entangling power of the integrable $\mathcal{\hat U}_0$ model with open boundary condition \cite{Mishra2015,Pal2018} is maximum at times where OTOC is maximum. This is consistent with the so-called OTOC-RE theorem at infinite temperature that related OTOC to the second Renyi entropy $S_V^2$ as $ C(n)\sim 1-e^{-S_V^2}= 1-\Tr \rho_V^2$ \cite{Hosur2016,FAN}, where $S_V^2=-\log \Tr_V(\rho_V^2)$ behaves like von Neumann entropy \cite{FAN,Bergamasco}. Here $\rho_V=\Tr_W[\rho]$ is the reduced density matrix for the partition scheme for the block operators defined in Fig.~\ref{block_operator}.

The exact vanishing of the OTOC at  $n=2mN$, $m\in\mathbb{Z}^+$, follows as it has been shown earlier that the quasienergies of the $\mathcal{\hat U}_0$ are in the multiples of  $\pi/(2N) $ such that as $\mathcal{\hat U}_0^{2  N}=I$ \cite{naik2019controlled}, therefore, in this case $\hat W(n=2mN)=\hat W$ and the commutator $[\hat W(n=2mN),V]$ becomes zero.

Similar to the integrable case, the nonintegrable $\mathcal{\hat U}_x$ case also shows a periodic behavior but the periodicity has a non-trivial unknown dependence on the system size [Fig. \ref{pi4_hx1_hz1_cf_nint_p0_N}(c)]. Again, the OTOC grows approximately quadratically ($b\approx2$) and independent of the system size [Fig. \ref{pi4_hx1_hz1_cf_nint_p0_N}(d)]. However, there are increasing fluctuations and the maximum value attained is only about 1.5 times the random matrix value of $C(\infty)$ and 
if there is any system size dependence, it is weak.

 Taking $\hat V$ and $\hat W$, as random matrices drawn from GUE, the power-law growth of OTOC gives way to initial exponential saturation in both integrable $\mathcal{\hat U}_0$ and  nonintegrable $\mathcal{\hat U}_x$ systems.  The exponent is nearly the same in both the cases ($\mu\approx0.77$ for $h_x=0)$ and $(\mu \approx0.85$ for $h_x=1$) as shown in Fig. \ref{pi4_hx1_hz1_cf_nint_p0_N}(f). The saturation value, although transient in the integrable case, is to a good approximation the random CUE value $C(\infty)$. 
For the integrable $\mathcal{\hat U}_0$ case, the periodic oscillation with time period equal to $2N$ remains as this is a property of the propagator. The OTOCs averaged over the random matrices $\hat V$ and $\hat W$ for  $\mathcal{\hat{U}}_0$ and $\mathcal{\hat{U}}_x$ systems are exactly same as OPEE $E_l[\mathcal{\hat{U}}_0]$ and $E_l[\mathcal{\hat{U}}_x]$, respectively (See Eq. (\ref{OPEE_OTOC})). 
  
 \par 
 For this special set of parameters, the spectrum of the Floquet operators, both integrable $\mathcal{\hat U}_0$ and nonintegrable $\mathcal{\hat U}_x$  are highly degenerate and we could not conclude the nature of distribution from the shape of NNSD. We observe that a small shift in $\tau$ from $\pi/4$ lifts this degeneracy. Therefore, it is useful to explore the proximity of $\tau=\pi/4$ by defining a small parameter (let's say, $\epsilon=\pi/50$) such that the natural behavior of NNSD and OTOC does not change by adding/subtracting $\epsilon$ to $\tau=\pi/4$. We explore not only NNSD but also OTOC at the proximity of $\tau=\pi/4$.
 
In the integrable $\mathcal{\hat U}_0$ system with $\tau=\pi/4-\epsilon$, we see OTOC deviates from the periodic behaviour at $\tau=\pi/4$. Though we still see maxima and minima of OTOC  near $n=(2m-1)N$ and $2mN$  for $m\in\mathbb{Z}^+$, respectively. We observe that smaller the $\epsilon$, sharper the maxima (minima) approaching to $n=(2m-1)N$ ($n=2mN$). 
[Fig. \ref{pi4e_hx0_hz1_cf_int_p0_N}(a)]. We again get a quadratic power-law growth ($b\approx 2$) at $\tau=\pi/4-\epsilon$  and the exponent is independent of the system size [Fig.~\ref{pi4e_hx0_hz1_cf_int_p0_N}(b)]. NNSD corresponding to this case displays  nearly Poisson statistics in the integrable $\mathcal{\hat U}_0$ system [Fig.~\ref{pi4e_hx0_hz1_cf_int_p0_N}(c)].

\par
On the other hand, OTOC in the nonintegrable $\mathcal{\hat U}_x$ system at $\tau=\pi/4-\epsilon$ show a different behaviour than that at $\pi/4$. 
There is no degeneracy in the spectrum, and the  NNSD shows Wigner-Dyson distribution [Fig.~\ref{pi4e_hx1_hz1_cf_nint_p0_N}(e)].
The OTOC grows till scrambling time and then saturates to the random matrix value of $C(\infty)$, [Fig. \ref{pi4e_hx1_hz1_cf_nint_p0_N}(a)]. The exact periodicity displayed at $\tau=\pi/4$ is not stable to perturbations. Although the growth of OTOC is again quadratic ($b\approx2$) and independent of the system size as well as shown in Fig. \ref{pi4e_hx1_hz1_cf_nint_p0_N}(b).

Replacing $\hat V$ and $\hat W$  by  pre-scrambled RBOs we get a similar behavior of OTOC as that at $\tau=\pi/4$, in the $\mathcal{\hat U}_x$ system.. However, in the $\mathcal{\hat U}_0$ system, the OTOC does not vanish at $n=2mN$. This is due to the parameter $\epsilon$ which, if tending towards zero, lead to coinciding $\tau=\pi/4-\epsilon$ case with $\tau=\pi/4$. Ideally OTOC for RBOs should also vanish at $n=2mN$ due to the same reason that $\hat W(n=2mN)=\hat W$  but with $\tau=\pi/4-\epsilon$, we skip the moment of vanishing OTOC at $2mN$ kicks and get a dip only [Fig.~\ref{pi4e_hx1_hz1_cf_nint_p0_N}(c)]. Again, we can confirm that OTOC averaged over the pre-scrambled RBOs is exactly same as OPEE  as given in Eq. (\ref{OPEE_OTOC}).
 
Fig.~\ref{pi4e_hx1_hz1_cf_nint_p0_N}(d)  displays the initial exponential saturation of OTOC with nearly equal exponent in both integrable $\mathcal{\hat U}_0$ and nonintegrable $\mathcal{\hat U}_x$ system ($\mu \approx 0.53$ for $\mathcal{\hat U}_0$ and $\mu \approx 0.56$ for $\mathcal{\hat U}_x$). 
 
\section{Conclusion} 
\label{conclusion}
In this manuscript, we study the growth and saturation behavior of OTOC in both integrable $\mathcal{\hat U}_0$ and nonintegrable $\mathcal{\hat U}_x$ systems. The OTOC is calculated for observables as blocks of spins each consisting of $N/2$ spins defined as SBOs. Initially, we calculated OTOC by using the SBOs for various time periods and analyzed the early time behavior and saturation behavior. Later, we used  RBOs to learn about the saturation region of the system. 

Growth of OTOC in both integrable $\mathcal{\hat U}_0$ and nonintegrable $\mathcal{\hat U}_x$ system shows  power-law for all Floquet periods in between $0<\tau <\pi/2$ except $\pi/4$. This finding for nonlocal  block-spin as observables are consistent with single-site localized observables or total spin observables studied previously in the literature. At kick interval $\tau=\pi/4$, the field terms do not change the state; therefore, OTOC remains constant, even for the nonlocal block observables.
\par
Later we take special parameters ($J_x=1$, $h_{z}=1$, and $h_{z}=0/1$ and $\tau=\pi/4$) and calculate the OTOC for the nonlocal SBOs. In the integrable system, we see a periodic trend and the period of oscillation is twice the system size. We also observe that the maxima/minima are those points where von Neumann entropy is also maxima/minima.  In the nonintegrable $\mathcal{\hat U}_x$ case, periodic behavior does not show a trivial dependence on the system size. For $\tau=\pi/4$, OTOC shows a quadratic power-law growth in the integrable $\mathcal{\hat U}_0$ system till $n=N-1$ kicks. We see a quadratic power law for the nonintegrable $\mathcal{\hat U}_x$ system as well. Large degeneracy at $\tau=\pi/4$ makes NNSD inconclusive whether it is Poisson or Wigner-Dyson type. In order to study the behavior approaching this Floquet period,  we take a slightly lesser value of $\tau=\pi/4-\pi/50$. At this $\tau$, NNSD is Poisson type in the integrable $\mathcal{\hat U}_0$ system and Wigner-Dyson type in the nonintegrable $\mathcal{\hat U}_x$ system.

We also studied the post-scrambling behaviour of  OTOC. 
In the nonintegrable $\mathcal{\hat U}_x$ system, OTOCs by using SBOs show the exponential behaviour that is consistent with random matrix theory. In the nonintegrable system, saturation behavior can not be exactly defined by using the SBOs; therefore, we consider pre-scrambled RBOs  and calculate OTOCs.  We are getting the exponential saturation of the OTOC  in all the cases which is consistent with the behavior previously observed for the operator entangling power.
\par
 In general, for a  bipartite system, averaging over pre-scrambled random Hermitian observables, drawn from GUE, OTOC is exactly same as the operator entanglement entropy.
\bibliographystyle{apsrev4-2}
\bibliography{LE_prb_p0}

\appendix
\begin{widetext}
\section{Calculation of post-scrambling OTOC using random $U$}
\label{appendix1}

 We calculate long time saturation values of OTOC for spin-block operators $\hat V$ and $\hat W$ are calculated by replacing the unitary operator $\hat U$ with random CUE of size $2^N$ and averaging over it. Two- and four-point correlation functions $C_2(n)$ and $C_4(n)$ are calculated as below: 
 \subsection{Calculation of $C_2(n)$ }
 Two point correlation ($C_2(n)$) averaged over random $ U$ drawn from CUE of size $2^N$ is given by
\beqa
\nonumber
\ovl{C_2(n)}^{U}&=&\frac{1}{d_A d_B}\ovl{\Tr(\hat W(n)^2\hat V^2)}^{U}.
\eeqa
Since time evolution of $\hat W$ is given by Heisenberg time evolution  as $\hat W(n)=\hat U(n)^{\dagger}\hat W\hat U(n)$. Hence, 
\beqa
\ovl{C_2(n)}^{U}&=&\frac{1}{d_A d_B}\ovl{\Tr(\hat U^{\dagger}\hat W^2\hat U\hat V^2)}^{U},\\
&=&\frac{1}{d_A d_B}\sum_{j=1}^{d}\ovl{\langle j|\hat U^{\dagger}\hat W^2\hat U\hat V^2)|j\rangle}^{U}, \\ \nonumber
&=&\frac{1}{d_A d_B}\sum_{j,k,l,m}\ovl{\langle j|\hat U^{\dagger}|k\rangle \langle k|\hat W^{2}|l\rangle\langle l|\hat U|m\rangle \langle m|\hat V^{2}|j\rangle}^{U},\\ \nonumber
&=&\frac{1}{d_A d_B}\sum_{j,k,l,m} \ovl{\hat U_{kj}^{*}\hat U_{lm}}^{U}\hat W^2_{kl}\hat V^2_{mj}, \\ \nonumber 
\eeqa
Since, $\ovl{\hat U_{kj}^{*}\hat U_{lm}}^{U}=\sum_{j,k,l,m} \delta_{kl}\delta_{jm} |\hat U_{kj}|^2$ and $|\hat U_{kj}|^2=\frac{1}{d}$
\beqa
\nonumber
\ovl{C_2(n)}^{U}&=&\frac{1}{d_A d_B}\frac{1}{d}\sum_{j,k,l,m} \delta_{kl}\delta_{jm}\hat W^2_{kl}\hat V^2_{mj},\\ \nonumber
&=&\frac{1}{d_A d_B}\frac{1}{d}\sum_{k,j} \hat W^2_{kk}\hat V^2_{jj},\\ \nonumber
&=&\frac{1}{d_Ad_B}\frac{1}{d} \Tr (\hat W^2) \Tr (\hat V^2). 
\eeqa
Since, $d_Ad_B=2^N$. Hence $C_2(n)$ will be 
\beqa
C_2(n)&=&\frac{1}{2^{2N}} \hat \Tr(\hat W^2)\hat \Tr(\hat V^2).
\eeqa
Since, block observables are localized spin block observables defined by Eq.~(\ref{Block}). Then  calculate $\Tr(\hat W^2)$ will be
\beqa
\label{trw}
\Tr  (\hat W^2)&=&\frac{4}{N^2}\Tr\Bigg(\sum_{l=1}^{\frac{N}{2}}(\hat \sigma_l^{x})^2+\sum_{l\neq l^{'}}\hat \sigma_l^x\hat \sigma_{l^{'}}^x\Bigg).
\eeqa
By using the properties of Pauli operator, square of Pauli operators are equal to identity matrix. Hence first term of Eq. \ref{trw} will be equal to $\frac{2}{N}2^N$. And second term, $\sum_{l\neq l^{'}}\hat \sigma_l^x\hat \sigma_{l^{'}}^x$ is equal to zero because Pauli observable follow the anti-commutation relation.
Hence, $C_2(n)$ for the spin block observables is
\beqa
\ovl{C_2(n)}^{U}=\frac{1}{2^{2N}}\frac{4}{N^2}2^{2N}=\frac{4}{N^2}.
\eeqa
 \subsection{Calculation of $C_4(n)$ }
Four-point correlator averaged over random $ U$ drawn from CUE of size $2^N$ is given by
\beqa
\nonumber
\ovl{C_4(n)}^{U} &=& \frac{1}{d_Ad_B}\ovl{\Tr(\hat W(n)\hat V \hat W(n)\hat V)}^{U},\\ \nonumber
&=&\frac{1}{d_Ad_B}\ovl{\Tr(\hat U^{\dagger}\hat W\hat U\hat V \hat U^{\dagger}\hat W\hat U\hat V)}^{U},\\ \nonumber
&=&\frac{1}{d_Ad_B}\sum_{i_1,i_2,\cdot,i_8} \ovl{\langle i_1|\hat U^{\dagger}|i_2\rangle \langle i_2|\hat W|i_3\rangle\langle i_3|\hat U|i_4\rangle \langle i_4|\hat V|i_5\rangle\langle i_5|\hat U^{\dagger}|i_6\rangle \langle i_6|\hat W|i_7\rangle\langle i_7|\hat U|i_8\rangle \langle i_8|\hat V|i_1\rangle}^{U},\\ \nonumber
&=&\frac{1}{d_Ad_B}\sum_{i_1,i_2\cdot i_8} \ovl{\hat U_{i_1,i_2}^{*}\hat U_{i_3,i_4}\hat U_{i_6,i_5}^{*}\hat U_{i_7,i_8}}^{U}\hat W_{i_2,i_3}\hat V_{i_4,i_5}\hat W_{i_6,i_7}\hat V_{i_8,i_1},\\ \nonumber
&=&\frac{1}{d_Ad_B}\sum_{i_1,i_2 \cdot i_8}\Bigg(\delta_{i_2,i_3}\delta_{i_1,i_4}\delta_{i_6,i_7}\delta_{i_5,i_8}|\hat U_{i_2,i_1}|^2|\hat U_{i_6,i_5}|^2 \hat W_{i_2,i_3} \hat V_{i_4,i_5} \hat W_{i_6,i_7} \hat V_{i_8,i_1}\\ \nonumber 
&+&\delta_{i_2,i_7}\delta_{i_1,i_8}\delta_{i_3,i_6}\delta_{i_4,i_5}|U_{i_2,i_1}|^2|\hat U_{i_3,i_4}|^2 \hat W_{i_2,i_3} \hat V_{i_4,i_5} \hat W_{i_6,i_7} \hat V_{i_8,i_1}\Bigg) \\ \nonumber
&-&\frac{1}{d_Ad_B}\sum_{i_1,i_2 \cdot i_8}\Bigg(\delta_{i_2,i_3}\delta_{i_1,i_4}\delta_{i_6,i_7}\delta_{i_5,i_8}\hat U^{*}_{i_2,i_1}\hat U_{i_2,i_4}\hat U^{*}_{i_6,i_5}\hat U_{i_6,i_8}\hat W_{i_2,i_3} \hat V_{i_4,i_5} \hat W_{i_6,i_7} \hat V_{i_8,i_1}\\ \nonumber
&+&\delta_{i_2,i_7}\delta_{i_1,i_8}\delta_{i_3,i_6}\delta_{i_4,i_5}U^{*}_{i_2,i_1}\hat U_{i_6,i_5}\hat U^{*}_{i_2,i_1}\hat U_{i_6,i_5}\hat W_{i_2,i_3} \hat V_{i_4,i_5} \hat W_{i_6,i_7} \hat V_{i_8,i_1}\Bigg),\\ \nonumber
&=&\frac{1}{d_Ad_B}\frac{1}{d^2-1}\sum_{i_1,i_2 \cdot i_8}\Bigg(\delta_{i_2,i_3}\delta_{i_1,i_4}\delta_{i_6,i_7}\delta_{i_5,i_8} \hat W_{i_2,i_3} \hat V_{i_4,i_5} \hat W_{i_6,i_7} \hat V_{i_8,i_1}+\delta_{i_2,i_7}\delta_{i_1,i_8}\delta_{i_3,i_6}\delta_{i_4,i_5} \hat W_{i_2,i_3} \hat V_{i_4,i_5} \hat W_{i_6,i_7} \hat V_{i_8,i_1}\Bigg) \\  \nonumber
&-&\frac{1}{d_Ad_B}\frac{1}{d(d^2-1)}\sum_{i_1,i_2 \cdot i_8}\Bigg(\delta_{i_2,i_3}\delta_{i_1,i_4}\delta_{i_6,i_7}\delta_{i_5,i_8} \hat W_{i_2,i_3} \hat V_{i_4,i_5} \hat W_{i_6,i_7} \hat V_{i_8,i_1}+\delta_{i_2,i_7}\delta_{i_1,i_8}\delta_{i_3,i_6}\delta_{i_4,i_5} \hat W_{i_2,i_3} \hat V_{i_4,i_5} \hat W_{i_6,i_7} \hat V_{i_8,i_1}\Bigg), \\ \nonumber
&=&\frac{1}{d_Ad_B}\frac{1}{d^2-1}\sum_{i_1,i_2 \cdot i_8} \Bigg( \hat W_{i_1,i_2} \hat V_{i_1,i_5} \hat W_{i_6,i_6} \hat V_{i_5,i_1} +\hat W_{i_2,i_3} \hat V_{i_4,i_4} \hat W_{i_3,i_2} \hat V_{i_1,i_1}\Bigg) \\ \nonumber
&-&\frac{1}{d_Ad_B}\frac{1}{d(d^2-1)}\sum_{i_1,i_2 \cdot i_8} \Bigg( \hat W_{i_2,i_2}\hat V_{i_4,i_4} \hat W_{i_6,i_6} \hat V_{i_8,i_8} + \hat W_{i_2,i_3} \hat V_{i_4,i_5} \hat W_{i_6,i_7} \hat V_{i_8,i_1} \Bigg), \\ \nonumber
&=&\frac{1}{d_Ad_B}\frac{1}{d^2-1}\Bigg((\Tr \hat W)^2 (\Tr \hat V)^2+(\Tr \hat W^2) (\Tr \hat V)^2\Bigg) \\ \nonumber
&-&\frac{1}{d_Ad_B}\frac{1}{d(d^2-1)}\Bigg(\Tr (\hat W^2) \Tr(\hat V^2)+(\Tr \hat W)^2 (\Tr \hat V)^2\Bigg)+O\Bigg(\frac{1}{d(d^2-1)}\Bigg).
\eeqa
Considering traceless observables such that $\Tr (\hat W)=0$ and $\Tr(\hat V)=0$, and $d_Ad_{B}=d$ we get 
\beqa
\ovl{C_4(n)}^{U}&=&-\frac{1}{d}\frac{1}{d(d^2-1)}(\Tr \hat W^2) (\Tr \hat V^2),~~~~~~~~~~~~~~ \\ \nonumber
&=&-\frac{1}{d^2(d^2-1)}(\Tr \hat W^2) (\Tr \hat V^2).
\eeqa
For traceless observables $C_2(n)$ will be
\beqa
\ovl{C_2(n)}^{U}=\frac{1}{d^2}\Tr (\hat W^2) \Tr (\hat V^2).
\eeqa
Hence, OTOC for the traceless observables will be 
\beqa
\ovl{C(n)}^{U}&=&\ovl{C_2(n)}^{U}-\ovl{C_4(n)}^{U}\nonumber \\ &=&\frac{1}{d^2}(\Tr \hat W^2) (\Tr \hat V^2)\Bigg(1+\frac{1}{d^2-1}\Bigg),\nonumber \\
&=&\frac{1}{d^2}(\Tr \hat W)^2 (\Tr \hat V)^2\frac{d^2}{d^2-1}, \nonumber \\ &=&\frac{1}{d^2-1}(\Tr \hat W)^2 (\Tr \hat V)^2, \nonumber \\
&=&\frac{1}{2^{2N}-1}, \nonumber \\
&\approx& \frac{1}{2^{2N}} 
\eeqa
 \end{widetext}
\end{document}